\documentclass[
notitlepage,
nofootinbib,
aps,
pra,
longbibliography,
superscriptaddress,
floatfix,
twocolumn,
10pt]{revtex4-2}

\usepackage{bm}
\usepackage{booktabs}

\usepackage{cancel}
\usepackage{amsmath}
\usepackage{amssymb}
\usepackage{amsfonts}
\usepackage{amsmath}
\usepackage{mathtools}
\usepackage{physics}
\usepackage{microtype}

\usepackage{graphicx}
\usepackage[caption=false, justification=centerlast]{subfig}
\usepackage{xcolor}
\usepackage[colorlinks=true,linkcolor=teal,citecolor=teal,urlcolor=teal]{hyperref}
\usepackage{multirow}
\usepackage{orcidlink}

\usepackage{tikz}
\usetikzlibrary{intersections}

\usepackage{pifont}
\newcommand{\cmark}{\color{green}{\ding{51}}}
\newcommand{\xmark}{\color{red}{\ding{55}}}

\usepackage{float}

\usepackage{tabularx}
\usepackage[normalem]{ulem}
\usepackage[ruled,lined]{algorithm2e}
\SetKw{Continue}{continue}

\usepackage{amsthm}
\newtheoremstyle{sltheorem}
{}                
{}                
{\slshape}        
{}                
{\bfseries}       
{.}               
{ }               
{}                
\theoremstyle{sltheorem}

\theoremstyle{definition}

\DeclarePairedDelimiter{\bbra}{\langle\!\langle}{|}
\DeclarePairedDelimiter{\kket}{|}{\rangle\!\rangle}
\newcommand{\ddyad}[1]{\kket{#1}\!\bbra{#1}}

\newcommand{\kketbbra}[2]{\kket{#1}\!\bbra{#2}}

\newcommand{\id}{\mathbb{I}}

\newcommand{\Var}[1]{\text{Var}\qty[#1]}

\DeclareMathOperator*{\argmin}{arg\,min}

\newcommand{\eg}{{\textit{e.g.}\ }}
\newcommand{\ie}{{\textit{i.e.}\ }}

\begin{document}

\title{Improving shadow estimation with locally-optimal dual frames}

\author{Keijo Korhonen\,\orcidlink{0009-0003-2647-3105}}
\email{keijo@algorithmiq.fi}
\affiliation{Algorithmiq Ltd, Kanavakatu 3C 00160 Helsinki, Finland}
\affiliation{QTF Centre of Excellence, Department of Physics, University of Helsinki, P.O. Box 43, FI-00014 Helsinki, Finland.}

\author{Stefano Mangini\,\orcidlink{0000-0002-0056-0660}}
\affiliation{Algorithmiq Ltd, Kanavakatu 3C 00160 Helsinki, Finland}
\affiliation{QTF Centre of Excellence, Department of Physics, University of Helsinki, P.O. Box 43, FI-00014 Helsinki, Finland.}

\author{Joonas Malmi\,\orcidlink{0000-0003-4916-3962}}
\affiliation{Algorithmiq Ltd, Kanavakatu 3C 00160 Helsinki, Finland}
\affiliation{QTF Centre of Excellence, Department of Physics, University of Helsinki, P.O. Box 43, FI-00014 Helsinki, Finland.}

\author{Hetta Vappula\,\orcidlink{0009-0006-6732-6744}}
\affiliation{Algorithmiq Ltd, Kanavakatu 3C 00160 Helsinki, Finland}

\author{Daniel Cavalcanti\,\orcidlink{0000-0002-2704-3049}}
\affiliation{Algorithmiq Ltd, Kanavakatu 3C 00160 Helsinki, Finland}


\begin{abstract}
Accurate estimation of observables in quantum systems is a central challenge in quantum information science, yet practical implementations are fundamentally constrained by the limited number of measurement shots. In this work we explore a variation of the classical shadows protocol in which the measurements are kept local while allowing the resulting classical shadows themselves to be correlated. By constructing locally optimal shadows, we obtain unbiased estimators that are competitive with state-of-the-art methods in terms of measurement overhead, while requiring only single-qubit measurements and allowing for the estimation of any observable in pure post-processing, reducing estimation errors by orders of magnitude over standard classical shadows. We validate our approach through numerical experiments on molecular Hamiltonians with up to 40 qubits consistently observing significant reductions in estimation errors, including for estimations of multiple chemically relevant observables simultaneously.
\end{abstract}

\maketitle


\section{Introduction}
One of the central challenges in quantum computation and information processing is the extraction of information from a quantum system. Although the most accurate strategy would be to measure directly in the eigenbasis of the target observable~\cite{MassarUncertainty2007}, this is generally infeasible: the diagonalization of complex observables is generally hard, and implementing the corresponding measurement is typically prohibitively difficult in practice. One possible solution is to decompose the observable in a product basis, \eg the Pauli basis, and estimate each element individually. A paradigmatic example of this measurement problem is that of estimating with good accuracy the energy and properties of quantum many-body systems prepared on quantum hardware, considered as one of the most promising candidates for near-term useful quantum advantage~\cite{georgescu2014quantum,mcardle2020quantum}. In this context, molecular Hamiltonians prove especially hard to measure, since the number of Pauli observables to be estimated scales badly with the system size $\smash{\mathcal{O}\qty(n^4)}$ with $n$ being the number of qubits~\cite{cao2019quantum}, thus becoming prohibitive already on systems of moderate size with an experimentally feasible amount of measurement shots.

Several approaches have been proposed in the literature to address this estimation problem. One way is to use the structure of the observable to devise observable-specific estimation protocols, an example being Pauli-grouping methods~\cite{WuOGM2023, Crawford2021PauliGrouping, yen2023deterministic, GreschShadowGrouping2025} that aim to reduce the measurement overhead by simultaneously measuring commuting terms. More recently, shadow tomography techniques and variants thereof have also emerged as promising candidates~\cite{HuangShadows2020, Paini2021estimating, Bertoni2023shallow, huang2021efficient, vankirk2024derandomized}, relying on the ``measure first, ask later'' paradigm where informationally-complete measurements are performed on the state, and subsequently the measurement outcomes are post-processed to estimate the property of interest.

In this context, relying on the equivalence between shadow techniques and well-known tomographic methods based on informationally-complete (IC) measurements~\cite{ScottTightICPOVM2006, DArianoICMeasurements2004, Perinotti2007optimalestimationensembleaverages, ShadowTomographyDualInnocenti2023, ZhouOCPOVM2014}, recent works have proposed improved randomized techniques that leverage previously unused degrees of freedom in the estimators to provide predictions with lower statistical errors~\cite{Malmi2024EnhancedEstimation, Fischer2024DualOptimization, CaprottiDualOptimisation, ManginiICE24}. However, these methods are either limited in terms of system sizes, can only account for local correlation in the post-processing, or rely on potentially hard iterative optimization routines.

In this work, we build upon a recently introduced method for computing low-variance unbiased estimators of any observable~\cite{Fischer2024DualOptimization}, making use of what we call correlated $k$-locally optimal ($k$-LO) classical shadows. These shadows are constructed from the measurement outcomes of an informationally-over-complete measurement in a way that they encode crucial information about the underlying state, while remaining classically efficient to manipulate. This state-aware property is in stark contrast with usual classical shadows, which instead rely on state estimators which are instead fully agnostic with respect to the measured quantum state and depend only on the measurement channel (\ie the measurement performed). In particular, the approach considered here makes use of the entanglement structure of the state by associating to each measurement outcome a shadow composed of correlated blocks of qubits.

\begin{figure*}[!ht]
    \centering
    \includegraphics[width=0.975\textwidth]{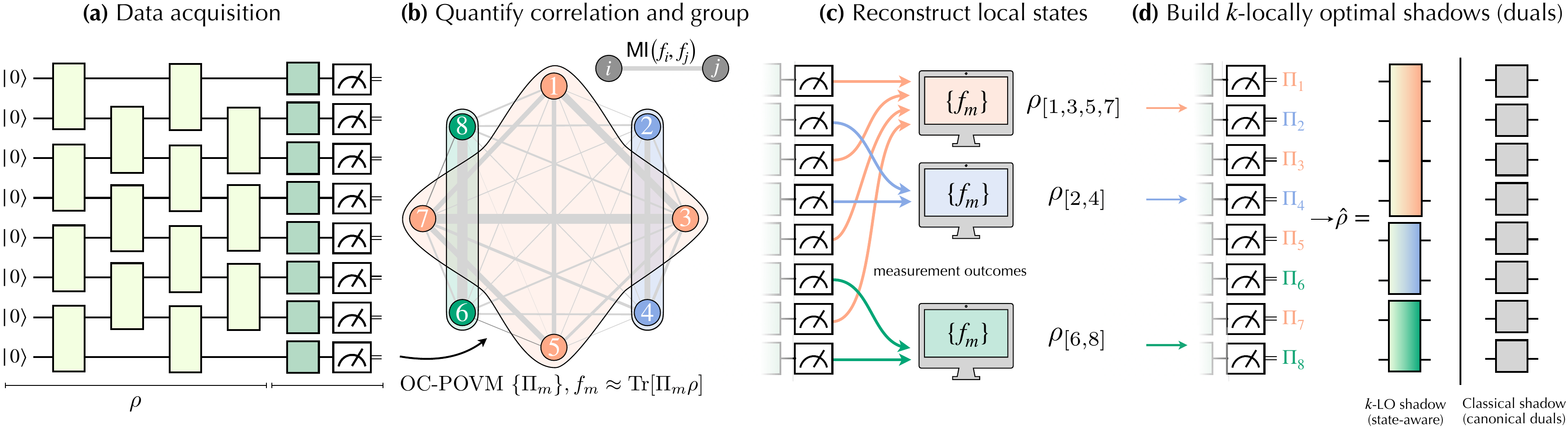}
    \caption{Summary of proposed methodology to construct state-aware correlated and locally-optimal shadows. \textbf{(a)} A quantum state $\rho$ is measured with an \textit{overcomplete} (OC) POVM, \eg single-qubit random Pauli measurements. \textbf{(b)} Using the observed frequencies, we build a graph of the pair-wise mutual information between the qubits, and partition it into disjoint groups of highest-correlated qubits, each of size at most $k$ ($k=4$ in the figure). \textbf{(c)} Reconstruct the local reduced density matrices for each of the groups from the measurement data using partial state tomography. \textbf{(d)} Construct optimal duals (shadows) for each of the groups. The corresponding global shadow is a tensor product of correlated $k$-qubit shadows which are locally-optimal for each the groups. Usual classical shadows are instead tensor products of single qubit operators and state-agnostic, in that they do not use available knowledge on the measured state.}
    \label{fig:main-image}
\end{figure*}

The main idea of the protocol is summarized in Fig.~\ref{fig:main-image} and works as follows. A quantum state is measured with an informationally over-complete measurement ---a common example being random single-qubit Pauli measurements---, with a given measurement budget. From the observed measurement frequencies, we construct a graph of the pair-wise mutual information between the qubits, and partition it so that highest-correlated qubits are grouped together, by imposing that a single group can contain at most $k$ qubits. We then perform partial state tomography for each of the identified groups, and then compute the corresponding optimal shadows~\cite{ShadowTomographyDualInnocenti2023}, which gives, by design, a classical representation of the local state that maximize the precision of the reconstruction. The final shadow is then given by joining together these $k$-locally optimal ($k$-LO) shadows constructed for each of the groups.

We tested this method extensively on several estimation tasks, including the estimation of the ground state energy of chemical Hamiltonians up to 16 qubits~\cite{hadfield2020github}, the estimation of multiple chemically relevant observables using the same measurement data on 16 qubits, and an application-driven example, the TLD1433 molecule up to 40 qubits, motivated by the Wellcome Leap ``Quantum for Bio'' program~\cite{wellcomeleapQ4BioProgram} as well as correlation functions in a 50-qubit spin model (Appendix~\ref{app:results-ising}). We observe that the $k$-LO shadows outperform standard classical shadows \cite{HuangShadows2020}, state-of-the-art approaches based on shadows techniques~\cite{HadfieldLBCS2022, HillmichGBCS-DD-2021}, and Pauli grouping approaches~\cite{WuOGM2023} whose single-shot variances are known, despite requiring either similar or fewer resources. In fact, while other methods rely on additional measurement circuits or are specific to measuring a single observable, the method exploited here is compatible with experimentally-friendly single-qubit measurements and can be used to estimate any observable with high accuracy with only classical post-processing.

The rest of the manuscript is structured as follows. In Sec.~\ref{sec:ic-estimation} we review the basics of informationally complete (IC) measurements, relation with shadow techniques, and introduce optimal duals (\ie shadows). In Sec.~\ref{sec:main-idea}, we describe the protocol for computing correlated locally-optimal shadows, discuss nuances of the method and compare it with other common state-of-the-art approaches. In Sec.~\ref{sec:methods}, we give technical details on each of the steps involved, including graph partitioning and partial state tomography. In Sec.~\ref{sec:results}, we show numerical evidence supporting our protocol and conclude in Sec.~\ref{sec:outlooks} with a discussion and interesting direction for future research.

\section{Observable estimation with informationally complete measurements}
\label{sec:ic-estimation}

Consider a measurement described by a Positive Operator-Valued Measure (POVM) with effects $\{\Pi_m\}$ satisfying $\Pi_m \geq 0$ and $\sum_m \Pi_m = \id$~\cite{Nielsen_Chuang_2010}. A measurement is called \textit{informationally complete} if the POVM effects span the space of linear operators $\mathcal{L}(\mathcal{H})$ acting on a Hilbert space $\mathcal{H}$~\cite{ScottTightICPOVM2006, DArianoICMeasurements2004}, i.e.
\begin{equation}
    \label{eq:ic-povm}
    O = \sum_{m} \omega_m \Pi_m\,,\quad \forall O \in \mathcal{L}(\mathcal{H})\,.
\end{equation}
In what follows we will refer to these POVMs as IC-POVMs. An IC-POVM can be seen as a frame to express any operator. In this context, duality theory says that for every IC-POVM, one can find a dual frame, composed by the set of dual operators $\qty{D_m}$ such that ~\cite{ShadowTomographyDualInnocenti2023, CasazzaIntroFiniteFrames2013}
\begin{equation}
    \label{eq:povm-duals-def}
    O = \sum_{m} \Tr[D_m O]\, \Pi_m = \sum_{m} \Tr[\Pi_m O] D_m\,,
\end{equation}
for every operator $O$. These two decomposition formulas come in handy for analyzing both observable estimation and state tomography tasks as Monte Carlo estimation procedures.

In fact, consider the task of estimating the expected value of an observable $O$ on a quantum state $\rho$. Let $o \in \mathbb{R}$ be a random variable taking values $\omega_m = \Tr[D_m O]$ each with probability $p_m = \Tr[\Pi_m \rho]$, where $p_m$ are probabilities of measuring each of the outcomes in the POVM. Then, using the first equality in Eq.~\eqref{eq:povm-duals-def}, one can estimate the desired expectation value $\expval{O} = \Tr[O \rho]$ by averaging the random variable $o$, namely
\begin{equation}
\begin{aligned}
    \label{eq:obs-estimation}
    \mathbb{E}_{o \sim (p, \omega)}[o] & = \sum_m p_m \omega_m = \sum_m \Tr[\Pi_m \rho] \Tr[D_m O] \\
    & = \Tr[O \rho]\,.
\end{aligned}
 \end{equation}
Similarly, applying the second equality in Eq.~\eqref{eq:povm-duals-def} to the state $\rho$, one can interpret the dual operators to the POVM effects as being a classical description of the post-measurement state. In fact, let $\sigma$ be a random operator taking values $\qty{D_m}$ each with probability $\qty{p_m}$, then its expectation value is
\begin{equation}
    \label{eq:state-tomography}
    \mathbb{E}_{\sigma \sim \qty(p, D)}[\sigma] = \sum_{m} p_m D_m = \sum_m \Tr[\Pi_m \rho] D_m = \rho\,.
\end{equation}

In practical scenarios one estimates the properties of quantum states from a limited number of experimental samples. Let $S$ be the number of shots acquired in an experiment. An unbiased estimator for the expectation value~\eqref{eq:obs-estimation} is then given by
\begin{equation}
\begin{aligned}
\label{eq:finite-stats-est}
    &\bar{o} = \sum_m f_m \omega_m  = \frac{1}{S}\sum_{s=1}^S \omega_{m_s},
\end{aligned}
\end{equation}
where $\bar{o}$ is the sample mean. The estimated variance is
\begin{equation}
\begin{aligned}
\label{eq:finite-stats-var}
    &\Var{o} = \sum_m f_m \omega_m^2 - \qty(\sum_m f_m \omega_m)^2, \\
    &\Var{\bar{o}} = \frac{\Var{o}}{S},
\end{aligned}
\end{equation}
where $\Var{o}$ is the sample variance and $\Var{\bar{o}}$ is the variance of the sample mean estimator. These quantities approach their infinite-statistics counterparts when $S \rightarrow \infty$, hence $f_m \rightarrow p_m$. By the central limit theorem, the sample mean estimator $\bar{o}$ converges to the true expectation value $\bar{o} \rightarrow \expval{O}$ as $\sqrt{\Var{o}/S}$.

It is then clear that, in addition to using more measurement shots $S$, one can obtain precise estimations by minimizing $\Var{o}$, that is finding coefficients $\omega_m=\Tr[D_m O]$ whose distribution has a small variance.

\subsection{Classical shadows and canonical dual frames}

The classical shadows protocol developed in Ref.~\cite{HuangShadows2020} is a particular case of this framework (see Ref.~\cite{ShadowTomographyDualInnocenti2023} for an extensive discussion on the topic), in which the IC measurement is implemented via randomized projective measurement and the classical shadows are a particular case of dual operators called \textit{canonical duals}, defined as
\begin{equation}
\begin{aligned}
    \label{eq:can-duals-def}
    &\kket{D^{\text{can}}_m} = F_{\text{can}}^{-1} \kket{\Pi_m}\,, \\
    &F_{\text{can}} \coloneqq \sum_{m=1}^{M} \frac{1}{\Tr[\Pi_m]}\ddyad{\Pi_m}\,,
\end{aligned}
\end{equation}
where we introduced the vectorized operators $\Pi_m\! \rightarrow\! \kket{\Pi_m}$ and $D_m \!\rightarrow \! \kket{D_m}$~\cite{WoodTNGraphCalculus} according to
\begin{equation}
    \label{eq:vectorised}
    O=\sum_{i,i} o_{i,j} \dyad{i}{j}\rightarrow \kket{O}=\sum_{i,j}o_{i,j}\ket{i,j}.
\end{equation}
With this notation, the decomposition formula relating effects and duals~\eqref{eq:povm-duals-def} can be rewritten more concisely as $\sum_m \kketbbra{D_m}{\Pi_m} = \id$. The (super)operator $F_{\text{can}}$ is called canonical \textit{frame operator}, and it is used to construct the duals starting from the measurement effects. Notably, the canonical frame operator plays the same role of the measurement channel in shadow tomography terminology~\cite{HuangShadows2020, ShadowTomographyDualInnocenti2023, ManginiICE24}.

\subsection{Optimal dual frames}
\label{sec:opt-duals}

Whenever the number of outcomes $m \in \qty{1, \ldots, d}$ is larger than the dimension of the space $d>\dim{\mathcal{L}(\mathcal{H})}$, the POVM is said to be \textit{overcomplete} (OC-POVM), and its duals are not uniquely defined~\cite{DArianoICMeasurements2004}. Furthermore, different choices of duals might provide estimators with different statistical properties~\eqref{eq:finite-stats-var}.

In fact, if one has perfect knowledge of the quantum state $\rho$ to be measured with the OC-POVM, then there is an optimal choice for the dual operators that minimize the statistical errors associated to the estimations. These \textit{optimal duals} are defined as~\cite{ZhouOCPOVM2014, ShadowTomographyDualInnocenti2023}
\begin{equation}
\begin{aligned}
    \label{eq:opt-duals-def}
    &\kket{D^{\text{opt}}_m} = F_{\text{opt}}^{-1} \kket{\Pi_m}\,, \\
    &F_{\text{opt}} \coloneqq \sum_{m=1}^{M} \frac{1}{\Tr[\Pi_m \rho]}\ddyad{\Pi_m}\,,
\end{aligned}
\end{equation}
which explicitly depend on the measurement outcomes probabilities $p_m = \Tr[\Pi_m \rho]$. These duals can be shown to minimize the expected mean squared error for state reconstruction $\mathbb{E}_{\sigma}[\norm{\sigma - \rho}^2_2]$ and, more importantly for our investigation, the estimation variance of \textit{any} observable estimation process~\eqref{eq:obs-estimation}. Formally, let $\Pi = \qty{\Pi_m}$ denote the POVM and $D = \qty{D_m}$ some dual operators to the POVM~\eqref{eq:povm-duals-def}, then the optimal duals~\eqref{eq:opt-duals-def} are such that
\begin{equation}
\begin{aligned}
    \label{eq:opt-duals-min-var}
    & \qty{D^{\text{opt}}_m} = \argmin_{D} \Var{O; \rho, \Pi, D} \\
    & \Var{O; \rho, \Pi, D} = \Var{o} \coloneq \mathbb{E}\qty[o^2]-\mathbb{E}[o]^2 \\
    & \phantom{\Var{O; \rho, \Pi, D}} = \sum_m p_m \omega_m^2 - \qty(\sum_m p_m \omega_m)^2
\end{aligned}
\end{equation}
with probabilities $p_m = \Tr[\Pi_m \rho]$, observable coefficients $\omega_m = \Tr[D_m O]$, and $\Var{O; \rho, \Pi, D}$ being the (infinite-statistics) variance associated to estimating observable $O$ on state $\rho$, using measurements $\Pi$ and duals $D$.

Importantly, the optimal duals defined in Eq.~\eqref{eq:opt-duals-def} yield the lowest possible variance compared to other duals for \textit{any} observable. In other words, for a fixed choice of IC measurement, the best post-processing of the measurement data is dictated only by the state, and not by the observables. We elaborate more on the differences between state- and observable-dependent measurement strategies in Sec.~\ref{ssec:comparison}.

\subsection{Practical limitations of optimal dual frames}
\label{ssec:limitations-opt-duals}
Despite their appealing statistical properties, optimal duals~\eqref{eq:opt-duals-def} cannot be used in practice for large systems. First of all, in order to compute them one would need perfect knowledge of the measured state, which is typically not available due to the difficulty of performing state tomography on multi-qubit systems. Second, even assuming that the state is known, for a system of $n$ qubits the frame operator is an exponentially large matrix in $n$, which makes obtaining and storing the optimal duals a challenging task.

One possible approach to address the first problem is to replace the measurement probabilities $p_m$ with observed frequencies $f_m$ in Eq.~\eqref{eq:opt-duals-def}. This approach has been already proposed in the literature and tested for small systems~\cite{ZhouOCPOVM2014, Fischer2024DualOptimization}. However, there are two drawbacks. First, for large systems the number of observed outcomes in a typical experiment tend to be just a tiny fraction of the total number of possible outcomes. Second, as shown in App.~\ref{app:comparison-tomography}, this simple approach tends to produce estimators that are \textit{biased} unless one uses independent datasets for computing the duals and performing estimations. Thus, this method either suffers from doubled measurement overhead and worse variances, or is practically limited to systems consisting of a few qubits.

As elaborated in Sec.~\ref{sec:main-idea}, in this work we propose to use a data-driven approach to find an effective and classically compact set of state-dependent duals that yields unbiased estimators with lower statistical error compared to many other state-of-the art approaches.

\section{Locally-optimal dual frames}

\label{sec:main-idea}
\begin{table*}[ht!]
\setlength{\tabcolsep}{5pt}
\renewcommand{\arraystretch}{1.1}
\setlength{\aboverulesep}{0pt}
\setlength{\belowrulesep}{0pt}
\begin{tabular}{c|cccc}
\toprule
\multirow{2}{*}{Method} & \multicolumn{3}{c}{Measurement}  & \multicolumn{1}{c}{Post-processing} \\
 & Informationally complete & Observable agnostic & Single-qubit measurements & State-aware \\
\hline
CS-Pauli~\cite{HuangShadows2020}  & \cmark & \cmark & \cmark & \xmark \\
LBCS{\footnote{LBCS is not necessarily informationally complete, but can be made so by putting a threshold in the bias~\cite{korhonen2025practical}. The method can also take into account information from chemical reference states, which is however different from being fully state-aware in this context.}}~\cite{HadfieldLBCS2022}      & \cmark & \xmark & \cmark & \xmark \\
GBCS~\cite{HillmichGBCS-DD-2021}  & \xmark & \xmark & \cmark & \xmark \\
Shallow~\cite{Bertoni2023shallow} & \cmark & \cmark & \xmark & \xmark \\
Derand~\cite{huang2021efficient}  & \xmark & \xmark & \cmark & \xmark \\
DSS~\cite{vankirk2024derandomized}& \xmark & \xmark & \xmark & \xmark \\
OGM~\cite{WuOGM2023}              & \xmark & \xmark & \cmark & \xmark \\
AEQuO{\footnote{AEQuO can optionally be restricted to use ``bitwise commuting'' cliques, which can be measured using single-qubit measurements. The method also involves an adaptive routine with multiple calls to the quantum computer.}}~\cite{shlosberg2023adaptive}& \xmark & \xmark & \xmark & \cmark \\
ShadowGrouping~\cite{GreschShadowGrouping2025} & \xmark & \xmark & \cmark & \xmark \\
\hline
$\bm{k}$-\textbf{LO duals}        & \cmark & \cmark & \cmark & \cmark \\
\toprule
\end{tabular}
\caption{Summary of the properties of some of the most common measurement techniques to estimate observable expectation values. ``CS-Pauli'' refers to standard classical shadows with single-qubit Pauli measurements~\cite{HuangShadows2020}, ``LBCS`` to locally-biased classical shadows~\cite{HadfieldLBCS2022}, ``GBCS'' to globally-biased classical shadows~\cite{HillmichGBCS-DD-2021}, ``Shallow'' refers to shallow shadows~\cite{Bertoni2023shallow, Hu2021ShallowShadow, Ippoliti2024classicalshadows}, ``Derand'' to derandomized shadows~\cite{huang2021efficient}, ``DSS'' to derandomized shallow shadows~\cite{vankirk2024derandomized}, ``OGM'' to overlapped grouping measurement~\cite{WuOGM2023}, ``AEQuO'' to Adaptive Estimator of Quantum Observables \cite{shlosberg2023adaptive}, ``ShadowGrouping'' to the method from Ref.~\cite{GreschShadowGrouping2025}. ``Informationally-complete'' means that the measurement is such that, provided enough samples, one can unambiguously reconstruct the state of interest, see Eq.~\eqref{eq:ic-povm}. Observable-agnostic means that the measurement strategy (\ie the POVM) does not depend on the observable to be measured, which often implies that the measurement is then informationally-complete (LBCS are an example of a technique which is observable-specific but still IC). ``Single-qubit measurement'' means that the measurement technique only requires additional single-qubit gates (\eg randomized Pauli measurements). ``State-aware'' post-processing means that the measurement results are analyzed in a way that is tailored to the measured state.}
\label{tab:comparison}
\end{table*}

As discussed in Sec.~\ref{sec:opt-duals}, for OC measurements the choice of the duals operators is not unique. This opens up the possibility of looking for duals that decrease the variance of the estimator, an idea initially explored in Refs.~\cite{CaprottiDualOptimisation, Malmi2024EnhancedEstimation, ManginiICE24} by means of optimization.

In Ref.~\cite{Fischer2024DualOptimization}, the authors proposed using duals with a tensor-product structure, where each block corresponds to duals that are optimal for a subset of qubits. In this work, we build on this idea and improve their findings in two directions. First, we propose an algorithm to determine the optimal tensor-product structure for the duals. Second, we employ local tomography to reconstruct the quantum state on each block, which is then used to construct the locally optimal duals. The first improvement reduces the variance of the final estimator, while the second avoids the biased features present in the original proposal of Ref.~\cite{Fischer2024DualOptimization} (see App.~\ref{app:comparison-tomography}). Overall, this sensibly extends the reach of the method, making it practical and effective for a large number of qubits and in the finite-statistics regime, as confirmed by the numerical results reported in Sec.~\ref{sec:results}.

In what follows, we provide a step-by-step overview of our approach, graphically depicted in Fig.~\ref{fig:main-image}, and then proceed analyzing each of them in detail in Sec.~\ref{sec:methods}:

\begin{enumerate}
    \item \textbf{Data acquisition---} collect $S$ measurement shots of the OC-POVM $\qty{\Pi_m}$ on the target state $\rho$. It corresponds to panel (a) in Fig.~\ref{fig:main-image}.
    \item \textbf{Quantifying correlations---} compute the classical mutual information of the outcomes' frequencies for each pair of qubits in the system. Construct the graph of the mutual information where each node represents a qubit and edges connecting them their pairwise mutual information. It corresponds to panel (b) in Fig.~\ref{fig:main-image}.
    \item \textbf{Defining groups of qubits---} partition the MI graph into disjoint groups of qubits so that qubits within the same group are highly-correlated with each other. Different groups can have different sizes, but the biggest one can contain at most $k$ qubits. It corresponds to panel (c) in Fig.~\ref{fig:main-image}.
    \item \textbf{Local tomography---} for each of group of qubits, use the $S$ measurement samples to construct the corresponding reduced density matrices (RDMs). It corresponds to panel (d) in Fig.~\ref{fig:main-image}.
    \item \textbf{Locally-optimal duals construction---} for each of the groups and corresponding RDMs, use Eq.~\eqref{eq:opt-duals-def} to construct the locally optimal duals for that group. Finally take the tensor product of these local duals as the final duals for the estimation. It corresponds to panel (d) in Fig.~\ref{fig:main-image}.
\end{enumerate}

For this procedure to hold, we are assuming that the measurement of step $1$ acts locally on each qubit, so that one can use Eq.~\eqref{eq:opt-duals-def} for the locally optimal duals on each group independently, by restricting the sum to the POVM effects acting only on the group of interest. It is also possible to use measurements acting jointly on more qubits, if one also makes sure that the duals are computed accordingly: that is, qubits measured by the joint measurement end up in the same group when computing the duals.

The intuition for why this could constitute a good approach is the following. Assume we have a state $\rho$ of $n$ qubits and we have found an informed partition of them into $L$ disjoint groups $\mathcal{G}=(G_1, \ldots, G_L)$ each of size at most $|G_\ell| \leq k$, according to steps (1-3) above. Consider the product state given by joining all marginal states for each of the groups, that is
\begin{equation}
    \label{eq:grouped-state}
    \rho_\mathcal{G} \coloneq \bigotimes_{\ell = 1}^{L} \Tr_{\bar{G_\ell}}[\rho]
\end{equation}
where $\bar{G_\ell}$ denotes all the qubits not in group $G_\ell$.
One can realize that the $k$-locally optimal ($k$-LO) duals constructed according to steps 4-5 are then essentially trying to reproduce the optimal duals for such coarse-grained $k$-local approximation $\rho_\mathcal{G}$ of the original state $\rho$ (see App.~\ref{app:k-lo-duals} for an explicit derivation).

Whenever the original state $\rho$ has a $k$-local product structure as the one in Eq.~\eqref{eq:grouped-state}, or it is close to it, then the $k$-locally optimal duals will be close to the truly optimal ones. If this is not the case, the marginalized state $\rho_\mathcal{G}$ will, by construction, still capture the relevant correlation in the state $\rho$. Then, the duals constructed with probabilities $\Tr[\rho_\mathcal{G} \Pi_m]$ won't be too different to the true ones obtained with $\Tr[\rho \Pi_m]$, and thus still perform well in practice.

In other words, by properly partitioning qubits into highly correlated groups, the corresponding locally-optimal duals will provide an informative and classically efficient representation of the original state $\rho$, thus providing better estimators compared to any other state-agnostic processing strategies, like classical shadows~\eqref{eq:can-duals-def}.

\subsection*{Comparison with other approaches}
\label{ssec:comparison}
As clear from the numerical results reported in Sec.~\ref{sec:results}, $k$-LO duals produce estimators which are unbiased, have generally low statistical error for any observable, and are tailored to the state that is measured. Additionally, the current method does not rely on the use of any additional measurement circuit ---neither shallow~\cite{Bertoni2023shallow, Hu2021ShallowShadow, Ippoliti2024classicalshadows} or deep~\cite{HuangShadows2020}---, as hardware-friendly single-qubit random Pauli measurement are already sufficient to provide good estimators.

This is at contrast with several state-of-the-art observable estimation techniques proposed in the literature, whose measurement schemes are, for example, dependent on the specific observable to be measured (and thus often not IC), or require adding an additional circuit to perform the measurement. In Tab.~\ref{tab:comparison} we summarize the properties of some of the most common observable estimation techniques proposed in the literature.

Remarkably, thanks to the use of informationally-complete measurements and a smart state-aware post-processing, $k$-LO duals yields estimates which have lower or comparable statistical error than other techniques, despite using the simplest measurement strategy. This underlines the importance of the post-processing step in any estimation procedure.

The method proposed in Ref.~\cite{Fischer2024DualOptimization}, on which our work builds upon, would satisfy the same conditions as our $k$-LO duals method in Tab.~\ref{tab:comparison}. However, unlike $k$-LO duals, it suffers from bias when the same dataset is used for dual construction and estimation as well as produces higher-variance duals even with independent datasets (see Section \ref{sec:constructing-frame-op} and Appendix~\ref{app:comparison-tomography}). Additionally, in this work we also propose concrete scalable strategies for computing groupings and demonstrate the efficacy of the pipeline on significantly larger systems.

\section{Methods}
\label{sec:methods}
In this section, we detail the methodology used to construct $k$-locally optimal ($k$-LO) duals outlined in Sec.~\ref{sec:main-idea}. After measuring the system, the first step consists of the identification of the highest correlating groups via the classical mutual information, followed by the reconstruction of the local reduced states and construction of the corresponding optimal duals for the identified partitions.

Let us consider a system $\rho$ of $n$ qubits, and a local POVM $\{ \Pi_{m_1}\otimes\Pi_{m_2}\cdots \otimes\Pi_{m_n}\}_{m_1,\ldots,m_n=1}^d$, where each qubit is associated to $d$ possible outcomes. For example, the POVM corresponding to single-qubit random Pauli measurement has $d=6$ outcomes per qubit, and thus a total of $6^n$ of effects. By performing $S$ measurement shots, we obtain the measurement statistics given by the experimental frequencies $f_{m_1, \ldots, m_n} = \#(m_1, \ldots, m_n)/S$ where $\#(m_1, \ldots, m_n)$ denotes the number of times the combination of outcomes $(m_1, \ldots, m_n)$ was observed.

\subsection{Group the qubits based on their correlations}
\label{sec:graph_partitioning}
The first step in the construction of the duals is to group the qubits according to their measurement statistics. To do so, we compute the classical mutual information~\cite{Nielsen_Chuang_2010} of the POVM outcomes among all pair of qubits in the system. That is, we compute the quantity
\begin{equation}
    \label{eq:cmi-freqs}
    I(i\!:\!j) \coloneqq \sum_{m_{i}=1}^d\sum_{m_{j}=1}^d f^{[i, j]}_{{m_i} m_{j}} \log \frac{f^{[i, j]}_{{m_i} m_{j}}}{f^{[i]}_{m_{i}}f^{[j]}_{m_{j}}}\,,
\end{equation}
where
\begin{equation}
\label{eq:marginal-freqs}
\begin{aligned}
    &f^{[i]}_{m_{i}} = \sum_{\substack{m_\ell =1 \\ \ell \neq i}}^d f_{m_1 m_2 \ldots m_n}\,, \quad\quad i=1, \ldots, n \\
    &f^{[i, j]}_{m_{i} m_{j}} = \sum_{\substack{m_\ell =1 \\ \ell \neq i, j}}^d f_{m_1 m_2 \ldots m_n}\,, \quad i, j = 1, \ldots, n
\end{aligned}\,.
\end{equation}
are the two-qubit and single-qubit marginal frequencies of the measurement outcomes.

Based on the pairwise mutual information between all pairs of qubits in the system, we group the qubits according to their correlations using the following procedure:

\begin{enumerate}
    \item Given the set of all qubits $Q = \qty{q_1, \ldots, q_n}$, begin by finding the two qubits $q_i$ and $q_j$ with the largest mutual information. Define group $G_1 = \qty{q_i, q_j}$ composed by these two qubits.
    \item For the remaining qubits $q \in Q \setminus G_1$, find the qubit $q_l$ with the largest mutual information with group $G_1$, obtained by generalizing Eq.~\eqref{eq:cmi-freqs} for the joint marginal outcomes of qubits in $G_1$ and the marginal frequencies of qubit $q_l$. Then add $q_l$ to group, thus obtaining $G_1 = \qty{q_i, q_j, q_l}$.
    \item Repeat step 2 by adding qubits to $G_1$ until it reaches a previously specified maximum group size $k$.
    \item Once group $G_1$ has reached the maximum size, create a new group, $G_2$, and repeat the steps above using the remaining qubits $Q \setminus G_1$. Continue until all qubits are allocated to a group.
\end{enumerate}

The final result of this procedure is a splitting of the qubits into $L$ disjoint groups, each containing at most $k$ qubits:
\begin{equation*}
\begin{aligned}
    &\mathcal{G} = (G_1, \ldots, G_L)\quad\text{with}\quad |G_\ell| \leq k\,,\\
    &G_\ell \subset [n],~ \bigcup_{G \in \mathcal{G}} G = [n],~ G_{\ell_1} \cap G_{\ell_2} = \varnothing.
\end{aligned}
\end{equation*}

In App.~\ref{app:partitioning-comparison} we discuss other approaches for grouping the qubits based on common graph-partitioning protocols~\cite{schlag2022high-quality,traag2019from}, and show that the procedure above tends to provide higher quality solutions compared to these approaches.


\subsection{Reconstructing local reduced density matrices}
\label{sec:approx_local_states}
The result of the previous step is a partition of the $n$ qubits into $\mathcal{G}=(G_1, \ldots, G_L)$ mutually exclusive groups. For each of them, we want to construct the corresponding reduced density matrices (RDMs) based on the available measurement data. We do this by employing a state tomography protocol based on semi-definite programming \cite{Skrzypczyk-Cavalcanti-book, cattaneo2023self-consistent}, consisting of solving
\begin{equation}
    \label{eq:sdp}
    \tilde{\rho} = \argmin_{\substack{\sigma >0 \\ \Tr[\sigma]=1}} \sum_{m}\abs{f_m - \Tr[\sigma \Pi_m]}
\end{equation}
where $f_m$ are the empirical frequencies associated to measurement effect $\Pi_m$. For a fixed group $G \in \mathcal{G}$, we then first compute the marginal frequencies for the qubits in such a group as in Eq.~\eqref{eq:marginal-freqs} and then solve the associated local SDP problem~\eqref{eq:sdp}, with the POVM being restricted to the relevant subspace.

After repeating the tomographic procedure for all the groups, we end up with a set of RDMs describing the state of each group in the partition:
\begin{equation*}
    \text{Groups } \mathcal{G}=(G_1, \ldots, G_L) \rightarrow \text{RDMs } \qty{\rho_{G_1}, \ldots, \rho_{G_L}}\,.
\end{equation*}

Although this manuscript focuses on tomographic reconstructions of the RDMs, our procedure works with any classically available representation of them. For instance, the RDMs can be obtained from a classical approximation of the measured state. Using such RDMs sidesteps any potential bias from overfitting to the measurement dataset when the same data is reused for estimation, though our numerical evidence indicates that our tomographic approach does not in fact suffer from this issue in practice (see App.~\ref{app:comparison-tomography}). Additionally, classically constructed RDMs are free of the statistical noise inherent to tomography, and thus, if the approximation is sufficiently accurate, they can yield duals of higher quality.

\subsection{Constructing \texorpdfstring{$k-$}-LO duals}
The final step in the procedure to compute the $k$-LO duals is to use the RDMs found above with the equation for optimal duals~\eqref{eq:opt-duals-def}. That is, for each group of qubits $G_i \in \mathcal{G}$ and corresponding local density matrix $\rho_{G_i}$, we build the locally-optimal duals as
\begin{equation}
\begin{aligned}
    \label{eq:k-lo-shadow}
    &\kket{D_{m_{G_i}}^{G_i}} = F_{G_i}^{-1}\kket{\Pi^{G_i}_{m_{G_i}}}\\
    &F_{\rho{G_i}} \coloneqq \sum_{m_{G_i}} \frac{1}{\Tr[\Pi^{G_i}_{m_{G_i}} \rho_{G_i}]}\ddyad{\Pi^{G_i}_{m_{G_i}}}\,,
\end{aligned}
\end{equation}
where $m_{G_i}$ is a multi-index labeling the qubits in $G_i$, and $\Pi^{G_i}_{m_{G_i}}$ denote the POVM effects acting only on the subspace consisting of the qubits in $G_i$.

In App.~\ref{app:comparison-tomography} we report results comparing the effect of the tomographic protocol on the resulting locally-optimal duals, including the method used in refs.~\cite{ZhouOCPOVM2014, Fischer2024DualOptimization} that uses directly the observed marginal frequencies in spite of the probabilities from a reconstructed local state in Eq~\eqref{eq:k-lo-shadow}. As already mentioned in Sec.~\ref{ssec:limitations-opt-duals}, this approach tends to provide estimates which are biased whenever the number of shots is low, which is not the case for our method.

Putting everything together, to the $s-$th measurement outcome $m^s = m_1^s m_2^s\ldots m_n^s$ in a dataset of $S$ shots, one associates the correlated $k$-locally optimal shadow
\begin{equation}
    \label{eq:final-opt-duals}
    m^s = m_1^s m_2^s\ldots m_n^s \rightarrow D_{m_s} = \bigotimes_{G \in \mathcal{G}} D^G_{m_{G}^s}\,,
\end{equation}
and corresponding coefficient $\omega_{m^s} = \Tr[O D_{m^s}]$, which can be used to the estimation of the expectation value and standard error of observable $O$ through ~\eqref{eq:finite-stats-est} and ~\eqref{eq:finite-stats-var}.

\subsection{Constructing the frame operator from finite data}
\label{sec:constructing-frame-op}
\begin{figure}[ht!]
    \centering
    \includegraphics[width=0.9\columnwidth]{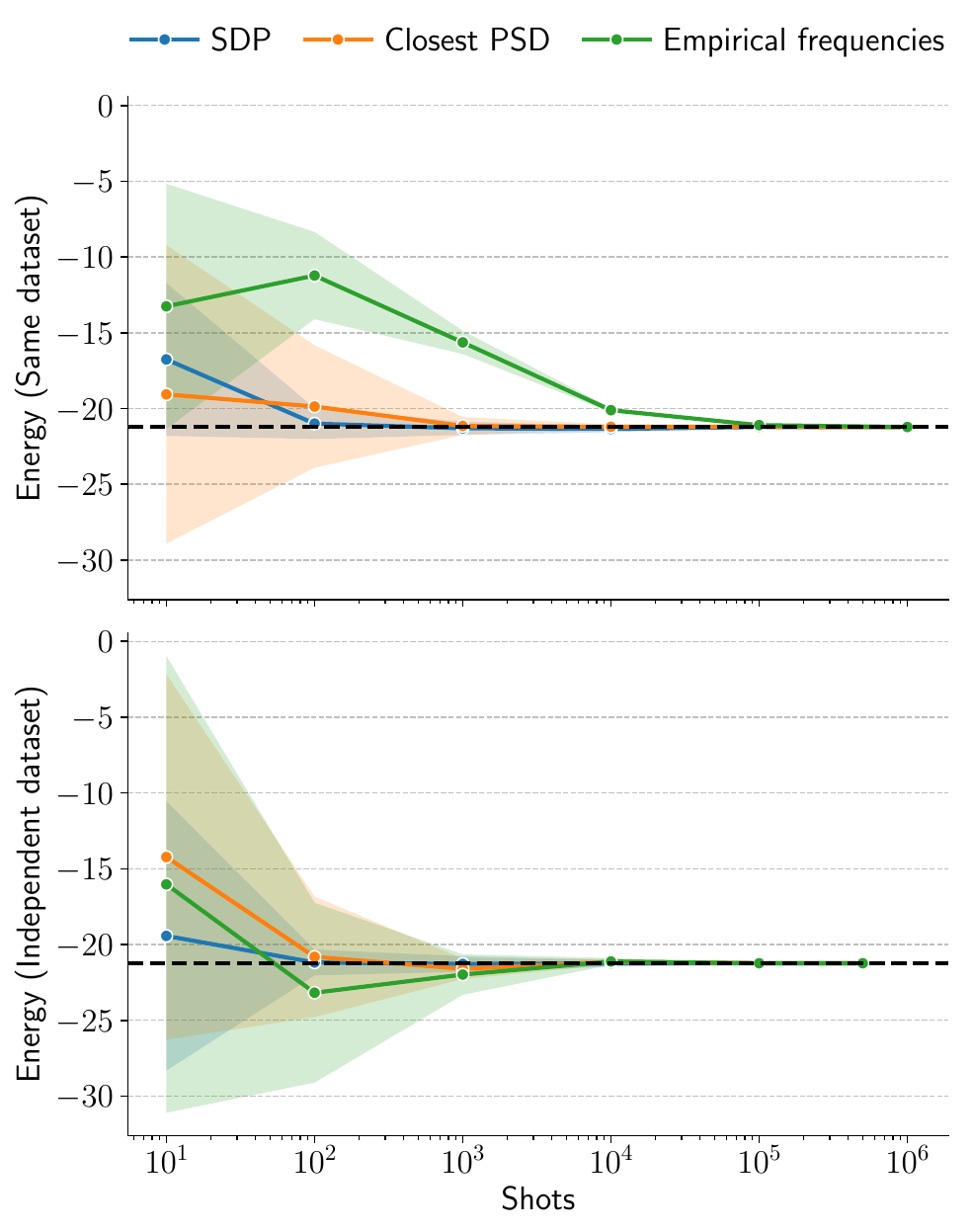}
    \caption{Comparison of methods to obtain the 4-LO frame operator for the ground state energy estimation of 16-qubit NH$_3$. The 4-LO duals and the energy estimations were computed using the same dataset in the top panel and with independent datasets in the bottom panel. For both cases ``SDP'' and ``Closest PSD'' methods produce unbiased estimators throughout, whereas the ``Empirical frequencies'' method (from Fischer et al.~\cite{Fischer2024DualOptimization}) is biased when the same data is used to construct the duals and perform the estimation. When an independent dataset is used instead, the estimation is unbiased but the variance is higher than the other methods. The method with empirical frequencies uses $S_{\textrm{bias}} = 6^4 = 1296$. The energies and their standard errors ($3 \sigma$) are shown in units of $\text{Ha}$.}
    \label{fig:rdm-method-comparison-nh3-both}
\end{figure}
Building the state-dependent frame operator in Eq.~\ref{eq:opt-duals-def} requires the outcome probabilities $\Tr[\Pi_m \rho]$, which are not available exactly at finite statistics.
A natural choice is to substitute the empirical frequencies $f_m$ directly, as in Refs.~\cite{ZhouOCPOVM2014, Fischer2024DualOptimization}, regularized by a constant $S_\text{bias}$ to avoid vanishing denominators.
Because the duals then depend on each $f_m$ through the $1/f_m$ terms of the frame operator, reusing the same dataset for dual construction and estimation couples the duals to the frequencies and biases the estimator at moderate shot counts.
We instead first reconstruct a local state from the data via the SDP~\cite{cattaneo2023self-consistent} or by projecting the unphysical approximation of the local states $\bar{\rho} = \sum_m{f_m D_m}$, where $D_m$ are canonical duals, onto the closest positive semi-definite (PSD) matrix~\cite{higham1988computing} and compute the probabilities from this estimate.
Mapping the $d^k$-dimensional frequency vector through the $4^k$-dimensional operator space acts as a bottleneck that attenuates the coupling between individual frequencies and the resulting duals, suppressing the bias to undetectable levels as shown in Fig.~\ref{fig:rdm-method-comparison-nh3-both}.
If any residual bias is a concern, an independent dataset for dual construction guarantees an unbiased estimator regardless of the reconstruction method.
We stress, however, that this is not merely a question of bias, as even when the empirical-frequency construction of Ref.~\cite{Fischer2024DualOptimization} is rendered unbiased by using independent datasets, it yields estimators with higher variance than ours, since the state reconstruction produces higher-quality duals.
The advantage of our approach is thus twofold: it removes the bias arising from data reuse, and it improves the performance of the duals themselves.
A full analysis is given in App.~\ref{app:comparison-tomography}.

\section{Results}
\label{sec:results}
\begin{figure*}[ht!]
    \centering
    \includegraphics[width=\textwidth]{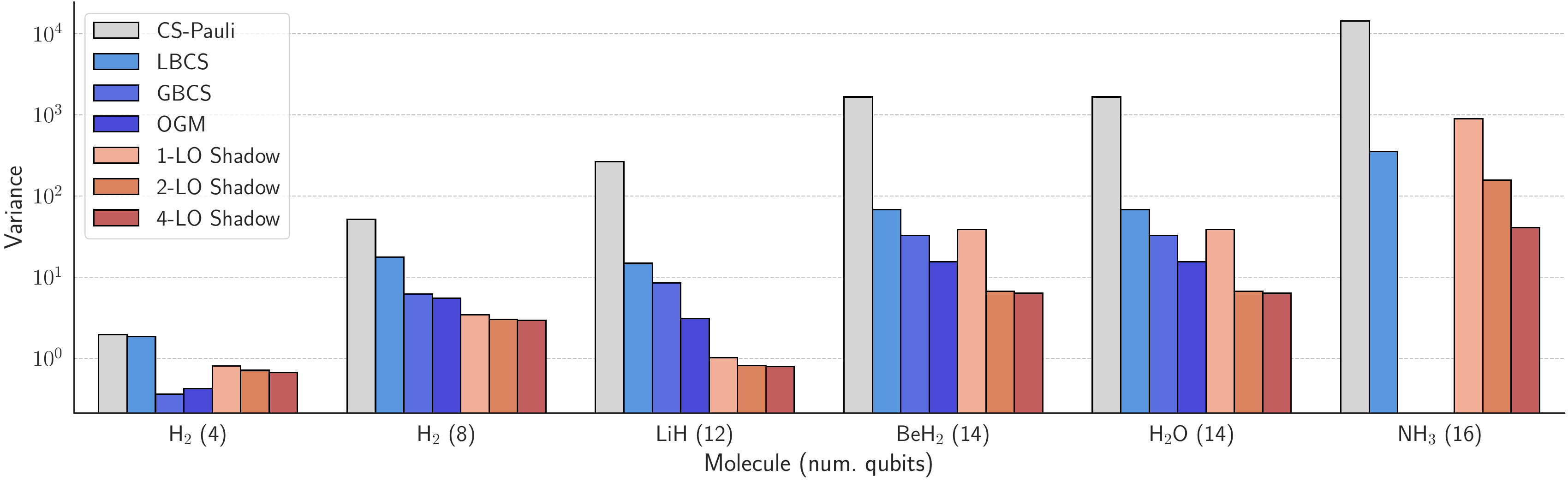}
    \caption{Comparison of $k$-locally optimal ($k$-LO) shadows against other methods (see Tab.~\ref{tab:comparison}) for the task of molecular energy estimation on a benchmark set of molecules (4 to 16 qubits), obtained via the Jordan--Wigner transformation from their fermionic Hamiltonians~\cite{hadfield2020github}. $k$-LO duals allow for observable estimators which are more precise than the state-of-the-art approaches considered, while using same or fewer resources. The variance is shown in units of $\text{Ha}^2$.}
    \label{fig:molecular-benchmark}
\end{figure*}

In this section, we show results from numerical simulations on the estimation of the ground and excited state energies of chemical molecules (Sec.~\ref{ssec:results-chem-en} and \ref{ssec:tld_numerics}), as well as other observables and properties of these states (Sec.~\ref{ssec:other_properties}). We show additional numerical results for the estimation of molecular energies including a comparison to values reported in literature in App.~\ref{app:variance-table}, and results for estimating the correlation functions in a Transverse Field Ising Model (TFIM) in App.~\ref{app:results-ising}.

In all the experiments reported below, the measurement is given by random single-qubit Pauli measurements, whose effects are
\begin{equation}
\begin{aligned}
    \label{eq:pauli-oc-povm}
    &\Pi_0 = \frac{\dyad{0}}{3}\,,\quad\Pi_2 = \frac{\dyad{+}}{3}\,,\quad \Pi_4 = \frac{\dyad{+i}}{3}\,, \\
    &\Pi_1 = \frac{\dyad{1}}{3}\,,\quad\Pi_3 = \frac{\dyad{-}}{3}\,,\quad \Pi_5 = \frac{\dyad{-i}}{3}\,.
\end{aligned}
\end{equation}
The multi-qubit POVM is given by tensor products of the local ones.

\subsection{Molecular energy estimation}
\label{ssec:results-chem-en}

We evaluate the performance of $k$-LO duals for molecular energy estimation on a standard benchmark set of molecules (4–16 qubits), obtained via the Jordan--Wigner transformation from their fermionic Hamiltonians. This set, publicly available in Ref.~\cite{hadfield2020github}, has been widely used in benchmarking measurement schemes~\cite{HadfieldLBCS2022, huang2021efficient, HillmichGBCS-DD-2021, WuOGM2023, vankirk2024derandomized, shlosberg2023adaptive}.

In Fig.~\ref{fig:molecular-benchmark}, we compare the variances of $k$-LO duals ($k \in \{1,2,4\}$) with other state-of-the-art methods whose exact single-shot variances are known (see Sec.~\ref{ssec:comparison}; numerical values in App.~\ref{app:variance-table}). The $k$-LO duals were computed using $S=10^6$ measurement shots, sufficient for high-quality 4-LO duals and feasible on current quantum hardware. The reported variance is the exact single-shot variance, obtained from the full probability distribution~\eqref{eq:opt-duals-min-var}. Except for the smallest 4-qubit example, 2- and 4-LO duals consistently achieve lower variances than the other methods. The 1-LO duals share the local structure of previous works~\cite{Malmi2024EnhancedEstimation, Fischer2024DualOptimization, CaprottiDualOptimisation}, differing only in their construction.

Furthermore, in App.~\ref{app:variance-table} we show that among single-qubit measurement schemes, $k$-LO duals exhibit the lowest root mean square error (RMSE).
We also show that $k$-LO duals remain competitive to correlated measurement schemes using entangling layers, such as AEQuO~\cite{shlosberg2023adaptive}. AEQuO consistently outperforms all of the methods we display, though at the cost of informational completeness, due to the observable-tailored nature of the method; an adaptive routine involving multiple calls to the quantum computer; and potentially gate-count in the form of additional two-qubit operations to implement an entangling measurement. See Sec.~\ref{ssec:comparison} for details.

\subsection{Towards energy estimation for large-scale molecules}
\label{ssec:tld_numerics}

\begin{figure}[ht!]
    \centering
    \includegraphics[width=\columnwidth]{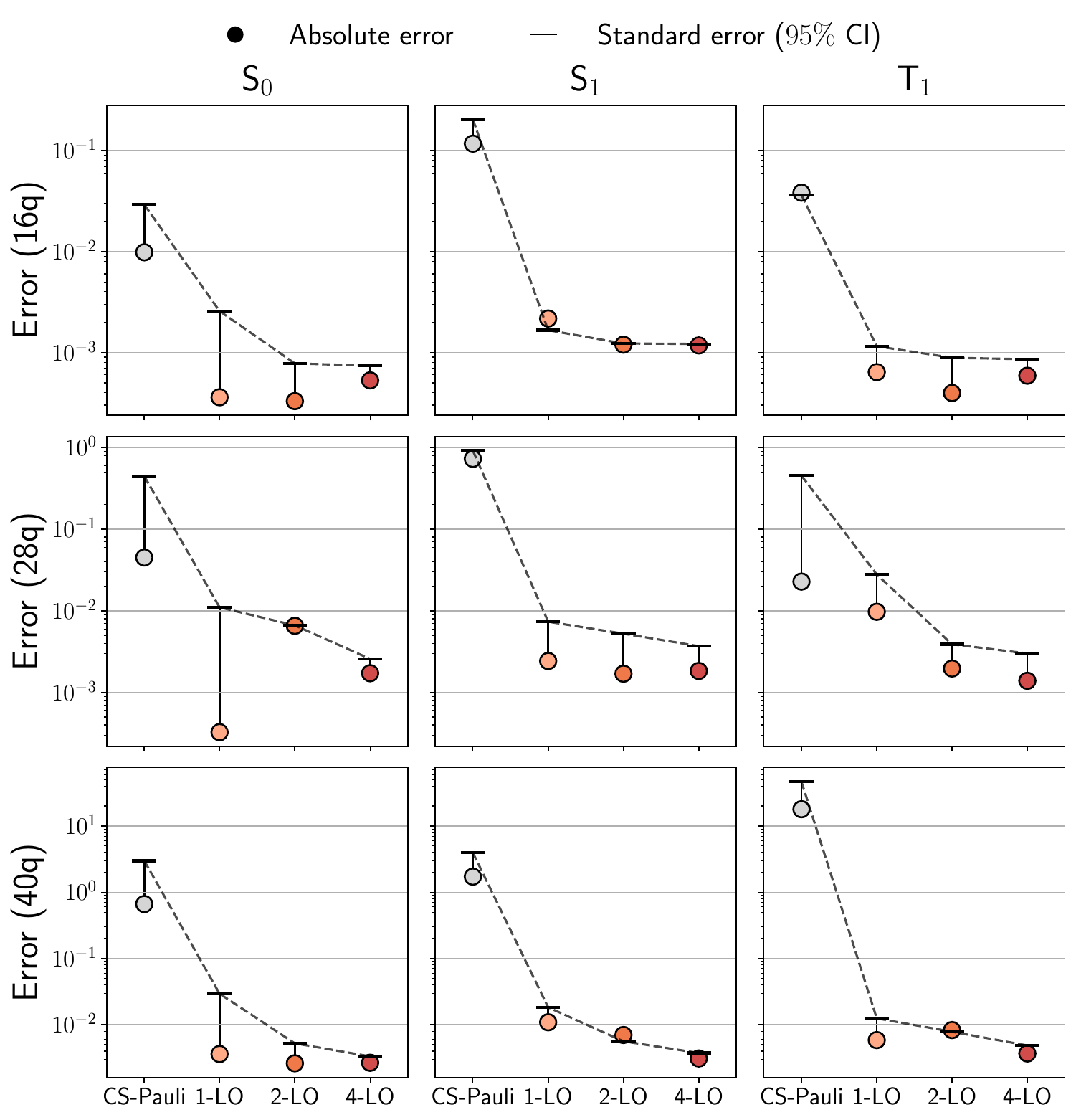}
    \caption{The errors of the estimation of various TLD1433 ansatz energies for the ground and excited states (columns) in different system sizes in terms of qubits (rows) using CS-Pauli and $k$-LO duals. The absolute error is computed with respect to the ansatz energy, and the standard error is computed using the data and our estimator. The errors are shown in units of $\text{Ha}$.}
    \label{fig:tld-benchmark}
\end{figure}

In the previous subsection, we analyzed $k$-LO duals for estimating ground state energies of molecular systems up to 16 qubits. Motivated by the Wellcome Leap ``Quantum For Bio" project~\cite{wellcomeleapQ4BioProgram}, we now consider an application-driven case, namely estimating energy gaps of TLD1433, a ruthenium-based photosensitizer used in photodynamic cancer therapy \cite{monro2018tld1433}. Specifically, we target precise estimations of the ground (S$_0$), first excited singlet (S$_1$), and triplet (T$_1$) state energies. We study TLD1433 across active spaces requiring $n=16$, $n=28$ and $n=40$ qubits, trading complexity for improved approximations of the ground and excited states of the complete molecule. For each size, ansatz circuits ---constructed via ADAPT-VQE~\cite{GrimsleyADAPT2019} and Majorana Propagation~\cite{miller2025simulation}--- achieve energy estimates within $10^{-2}$ accuracy.

In Fig.~\ref{fig:tld-benchmark}, we compare absolute and standard errors for CS-Pauli and $k$-LO duals. The absolute error is defined as $|\bar{o}- \Tr[\rho O]|$, while the standard error is defined as $\bar{\sigma}_{E}=\sqrt{\Var{\bar{o}}}$ (Eq.~\eqref{eq:finite-stats-var}), shown with 95\% confidence intervals, namely $1.96\cdot\bar{\sigma}_{E}$. Each plot corresponds to one state and system size, each estimation using the same dataset of $S=10^6$ shots sampled from uniform random Pauli measurements on a truncated MPS representation of the ansatz circuit, with limited bond dimension $\chi \le 50$~\cite{Schollwok2011DMRGMPS}.

The results show that $k$-LO duals reduce errors by orders of magnitude compared to CS-Pauli across all sizes, improving both accuracy and precision. Remarkably, all the CS-Pauli and the $k$-LO duals estimations corresponding to the same energy are performed using the same dataset, which shows that the variance obtained with CS-Pauli can be decreased significantly and easily in post-processing using $k$-LO duals. We also note that, while the comparison between the methods is fair, the MPS truncation performed on the state can limit the conclusions we can take from these results, as reduced entanglement in the truncated MPS may favor local measurements and $k$-LO duals. For instance, the magnitudes of the errors we report may differ from those on untruncated states.

\subsection{Estimation of multiple observables using the same IC data}
\label{ssec:other_properties}
One of the main advantages of shadow estimation approaches is their ability to estimate multiple observables in pure post-processing using the same data.
This is in contrast to derandomized or observable-specific protocols, which are often not informationally complete, and thus are unable to estimate observables besides the one that they were used for.

To assess the performance of the $k$-LO dual estimator in tasks beyond Hamiltonian estimation, we evaluate several observables that are central in molecular electronic structure: the number operator $N$, the total spin operator $S^2$ and the components of the spin angular momentum vector, $S_x$, $S_y$ and $S_z$.
These quantities are routinely used to diagnose particle‑number leakage, spin contamination, and symmetry breaking in variational and strongly‑correlated electronic states.
In particular, $S^2$ is a high‑weight, chemically meaningful observable that is difficult to estimate accurately using standard classical shadows.
In Table~\ref{table:other-observables} we show results of the estimation the aforementioned observables on the 16-qubit NH$_3$ molecule using CS-Pauli and $k$-LO duals.
Across different post-processing techniques, CS-Pauli and $k$-LO duals, the estimations were performed using the same $10^6$-shot dataset.

The results show that $k$-LO duals consistently retain the orders-of-magnitude reduction in the variance compared to CS-Pauli across all considered observables. This improvement is particularly significant for the high-weight observable $S^2$, where the error bars in the CS-Pauli estimate are too large to extract any meaningful information, while $k$-LO duals produce estimates whose error bars are small enough to reveal the underlying spin structure of the state in this shot regime.
It is worth stressing again that the $k$-LO duals do not depend on the target observable but only on the state and the measurement distribution. In the limit of $k$ approaching the system size $k = n$, access to the optimal dual frame implies that optimal estimation of any observable coincides with optimal state estimation with the OC-POVM~\cite{ShadowTomographyDualInnocenti2023}.
Our locally‑optimal construction works in an intermediate regime, by bringing the duals closer to the optimal ones, and therefore improving the estimation of all observables simultaneously, without any the need of any observable‑specific modifications.

\begin{table}[ht]
\centering
\begin{tabular}{l l r}
\toprule
Observable & Estimator & Estimate \\
\midrule
$H$ & Exact & $-66.8813$ \\
 & CS-Pauli & $-66.9291 \pm 0.1446$ \\
 & 1-LO & $-66.8712 \pm 0.0109$ \\
 & 2-LO & $-66.8747 \pm 0.0099$ \\
 & 4-LO & $-66.8786 \pm 0.0060$ \\
\midrule
$N$ & Exact & $10$ \\
 & CS-Pauli & $9.9970 \pm 0.0028$ \\
 & 1-LO & $9.9999 \pm 0.0006$ \\
 & 2-LO & $10.0000 \pm 0.0005$ \\
 & 4-LO & $10.0001 \pm 0.0005$ \\
\midrule
$S^2$ & Exact & $0$ \\
 & CS-Pauli & $0.4996 \pm 1.3064$ \\
 & 1-LO & $0.0394 \pm 0.1050$ \\
 & 2-LO & $-0.1259 \pm 0.1071$ \\
 & 4-LO & $-0.0346 \pm 0.0469$ \\
\midrule
$S_x$ & Exact & $0$ \\
 & CS-Pauli & $-0.0049 \pm 0.1391$ \\
 & 1-LO & $0.0001 \pm 0.0076$ \\
 & 2-LO & $0.0024 \pm 0.0078$ \\
 & 4-LO & $0.0009 \pm 0.0074$ \\
\midrule
$S_y$ & Exact & $0$ \\
 & CS-Pauli & $0.0049 \pm 0.1382$ \\
 & 1-LO & $0.0085 \pm 0.0090$ \\
 & 2-LO & $0.0004 \pm 0.0080$ \\
 & 4-LO & $-0.0048 \pm 0.0077$ \\
\midrule
$S_z$ & Exact & $0$ \\
 & CS-Pauli & $-0.0013 \pm 0.0014$ \\
 & 1-LO & $0.0006 \pm 0.0003$ \\
 & 2-LO & $0.0004 \pm 0.0003$ \\
 & 4-LO & $0.0003 \pm 0.0002$ \\
\bottomrule
\end{tabular}
\caption{Observable estimation results for the NH$_3$ (16 qubits) ground state using the same $10^6$-shot dataset from uniform Pauli measurements. The energies (\ie~observable $H$) and their errors are shown in units of $\text{Ha}$.}
\label{table:other-observables}
\end{table}

\section{Final remarks}
\label{sec:outlooks}
In this manuscript we have explored $k$-locally optimal ($k$-LO) duals, a state-dependent, correlated and locally optimal version of classical shadows that significantly improves the accuracy of any observable estimation task. Compared to standard classical shadows~\cite{HuangShadows2020, Bertoni2023shallow}, whose post-processing strategy only depends on the chosen measurement scheme, the $k$-LO shadows are tailored to the state being measured, thereby obtaining estimators with better statistical properties.

The method works by first measuring the state for a given number of shots with informationally-overcomplete POVMs ---a common example being random Pauli measurements---, and then using this data to partition the qubits into highly-correlated disjoint groups based on their pair-wise mutual information. Then, for each of the identified groups, we perform partial state tomography and explicitly build the corresponding locally-optimal duals (shadows) using known results in IC-POVM estimation theory~\cite{ShadowTomographyDualInnocenti2023, ZhouOCPOVM2014, OptimalProcessingDariano2007, ScottTightICPOVM2006}.

We have extensively tested the proposed methodology on several examples, including estimating the energy of molecular systems up to $n=16$ qubits, the estimation of multiple chemical observables on a molecular system of $n=16$ qubits and the TLD1433 molecule on $n=16,28,40$ qubits. In all these examples, we have seen that $k$-LO duals provide estimates that are competitive with or improve upon other state-of-the-art approaches, while using either comparable or fewer resources, in that they do not rely on knowledge of the observable to be measured, and are readily compatible with single-qubit random measurements. Overall, the results highlight the importance of the post-processing strategy in any estimation task, and especially the benefit of using knowledge of the state being measured.

As a subject for future studies, it would be interesting to have a more detailed and theoretical analysis of the statistical properties of the proposed estimators, and thus have a better understanding of the limiting cases where this approach may fail (see the discussion in App.~\ref{app:k-lo-duals}). We also note that our method is, in principle, compatible with any informationally complete POVM scheme consisting of purely local effects, such as LBCS \cite{HadfieldLBCS2022}. The combination of a more tailored, local POVM scheme, instead of uniformly sampled Pauli measurements, and $k$-LO duals would be insightful to study. Additionally, a natural future step would be to study the effect of combining this state-dependent property of the $k$-LO duals with other potentially useful techniques, like using correlated measurement schemes or observable-specific ones.

\section*{Acknowledgements}
Work on ``Quantum Computing for Photon-Drug Interactions in Cancer Prevention and Treatment'' is supported by Wellcome Leap as part of the Q4Bio Program. We thank the State preparation and Chemistry teams at Algorithmiq for providing the TLD1433 Hamiltonian and ansatz circuits.

\section*{Competing Interests}
Elements of this work are included in patent applications filed by Algorithmiq Oy currently pending with the European Patent Office.

\section*{Author Contributions}
The algorithm was conceived by KK, SM, and DC. KK, SM, HV, and JM contributed to the development and benchmarking of the code. The manuscript was written by KK, SM, and DC.


%

\newpage

\appendix
\onecolumngrid

\section{Remarks on the optimality of \texorpdfstring{$k-$}-LO duals}
\label{app:k-lo-duals}

Despite the physically motivated intuition behind $k$-LO duals explained in Sec.~\ref{sec:main-idea}, the method remains a heuristic one and it is not possible to prove that, in general, it will provide estimator with better statistical performances that standard classical shadows (canonical duals).

This is because the performance of any estimator highly depends on the specific state and observable under investigation and varies on a case-by-case scenario, thus hindering the possibility that a general optimality proof can be derived. In fact, as we show in the rest of this appendix with two toy examples, one can find pathological cases where $k$-LO duals can actually perform worse than canonical ones.

On a similar note, it is also difficult to prove that $k$-LO duals with larger locality $k_1$ are \text{strictly} better than those with lower locality $k_2<k_1$, even though this is always observed in numerical results, Sec.~\ref{sec:results}. One reason is because the groups are chosen with an heuristic approach which strongly depends on the specific state under consideration. Another reason is that, while it seems reasonable, to the best of our knowledge there is no proof that the more you coarse-grain a multi-qubit state, the further away you move from it. That is, given a multi-qubit system and two different partitions $\mathcal{G}_1$ and $\mathcal{G}_2$ of the qubits, with $\mathcal{G}_1$ containing larger groups, the marginalized product states obey $D(\rho, \rho_{\mathcal{G}_1}) \leq D(\rho, \rho_{\mathcal{G}_2})$, with $\rho_\mathcal{G}$ defined as in Eq.~\eqref{eq:grouped-state}, and $D(\cdot,\!\cdot)$ being a distance measure between quantum states. Indeed, the result not only depends on the distance measure used, but it is not even clear which measure is the most appropriate one in terms of optimality of the corresponding duals.

Overall, even without a general proof optimality, the physical intuition behind $k$-LO duals the method is strongly supported by the numerical results in Sec.~\ref{sec:results}, in that they outperform other state-of-the-art approaches for observable estimation tasks using similar or fewer resources.

In the rest of the appendix we review some basic concepts and definitions about duals operators to an OC-POVM, including various types of duals and notions of optimality. After these necessary introductions, we show in Sec.~\ref{sec:toy-examples} with two toy examples the differences between these various duals, and argue that in some pathological cases $k$-LO duals can perform worse than standard classical shadows.

\subsection{Optimal duals}
Consider a state $\rho$ and an OC-POVM with effects $\Pi = \qty{\Pi_m}$, each occurring with a measurement probability $p_m = \Tr[\rho \Pi_m]$. Also consider a set of duals $D = \qty{D_m}$ to the POVM effects, that is a set of operators that satisfies the reconstruction formula
\begin{equation}
    \label{eq:app_duals_def}
    O = \sum_m \Tr[D_m O] \Pi_m\,,~\forall O\,.
\end{equation}
Let $\mathcal{D}$ denote the set of all possible valid duals, \ie the set of all operators $D=\qty{D_m}$ that satisfy Eq.~\eqref{eq:app_duals_def}. The reconstruction precision of a given choice of duals $D$ can be measured with the state estimation mean squared error (MSE)~\cite{ShadowTomographyDualInnocenti2023, ZhouOCPOVM2014, Perinotti2007optimalestimationensembleaverages}, defined as
\begin{equation}
    \label{eq:mse_state}
    \mathcal{E}_{\rho}(D) \coloneqq \mathbb{E}_{\sigma \sim (p_m,\, D_m)} \qty[\norm{\sigma - \rho}^2_2] = \sum_m p_m \Tr[D_m^2] - \Tr[\rho^2]\,.
\end{equation}
with $\norm{X}^2_2 \coloneqq \Tr[X^\dagger X]$ being the Frobenius norm. The so-called \textit{optimal duals} are those that minimize the MSE, that is
\begin{equation}
    \hat{D} = \argmin_{D \in \mathcal{D}} \mathcal{E}_{\rho}(D)\,.
\end{equation}
In particular, as extensively discussed in Ref.~\cite{ShadowTomographyDualInnocenti2023}, these state-dependent optimal duals are the best ones in terms of precision for both state tomography and observable estimation tasks, irrespective of the observable.

\subsection{Optimal \texorpdfstring{$k-$}-local duals}
Let us now restrict ourselves to duals with a $k$-local structure, that is we consider duals that can be written as tensor products of operators each acting at most on $k$ qubits. Specifically, let $n$ be the number of sites in $\rho$, consider a partition $G = (g_1, \ldots, g_{\abs{G}})$ of these sites into $\abs{G}$ disjoint subsets each containing at most $|g_i| \leq k$ qubits and $\sum_{g \in G}|g|=n$. Consider dual operators that factorize over the groups
\begin{equation}
    \label{eq:duals-k-structured}
    D_m = \bigotimes_{g \in G}{D_{m_g}^{(g)}}\,,
\end{equation}
where $D^{(g)}_{m_g}$ is an operator acting on qubits in group $g$, and it is a matrix of size at most $\mathbb{C}^{2^k} \times \mathbb{C}^{2^k}$. Let $\mathcal{D}_G$ denote the set of duals that admit such $k$-local structure with partition $G$~\eqref{eq:duals-k-structured}. Given a state $\rho$, the best $k$-structured duals are those that minimize the MSE~\eqref{eq:mse_state}, namely
\begin{equation}
    \label{eq:k-struct-duals}
    \hat{D}_{G} = \argmin_{D \in \mathcal{D}_G} \mathcal{E}_{\rho}(D)\,.
\end{equation}
Importantly, since the minimization of the function above requires knowledge of the exponentially-many measurement outcomes probabilities $p_m = \Tr[\rho \Pi_m]$, these duals are not available in general for large number of qubits.

\subsection{\texorpdfstring{$k-$}-locally optimal (\texorpdfstring{$k-$}-LO) duals}
Finally, let us focus on the type of duals that we use in the main text, namely the $k$-LO duals introduced in Sec.~\ref{sec:methods}. For these, in addition to only considering duals having a $k$-local structure as in Eq.~\eqref{eq:duals-k-structured}, we also restrict ourselves to using only local information on the state $\rho$ under investigation.

Given a $k$-local partition $G=(g_1, \ldots, g_{\abs{G}})$ as above, we consider the reduced states $\rho_{g} = \Tr_{\bar{g}}[\rho],~g\in G$, where $\bar{g}$ is the complement of the subset $g$. For each of these local reduced states $\rho_g$, the corresponding locally optimal duals are those that minimize the associated MSE
\begin{equation}
\label{eq:mse_local}
    \hat{D}_g = \argmin_{D\in \mathcal{D}_g} \mathcal{E}_{\rho_g}(D)\,,\quad
    \mathcal{E}_{\rho_g}(D) = \sum_{m_g} \Tr[\rho_g \Pi^{(g)}_{m_g}] \Tr[D_m^2] - \Tr[\rho_g^2]\,,\quad \forall g \in G=(g_1, g_2, \ldots)\,,
\end{equation}
where $\mathcal{D}_g$ denote the set of duals for the space of qubits in group $g$, and we have also assumed that the global POVM $\Pi=\qty{\Pi_m}$ also admits a factorization over the groups $\smash{\Pi_m = \Pi^{(g_1)}_{m_{g_1}} \otimes \Pi^{(g_2)}_{m_{g_2}} \otimes \cdots }$. Also, let $\rho_G$ be the product state constructed by the reduced states of $\rho$ according to partition $G$, that is
\begin{equation}
    \label{eq:app_rho_part}
    \rho_G \coloneqq \bigotimes_{g \in G} \Tr_{\bar{g}}[\rho] = \bigotimes_{g \in G} \rho_g\,.
\end{equation}
Since everything has a product structure, one can check that the locally optimal duals in Eq.~\eqref{eq:mse_local} are those that minimize the MSE associated to such partitioned state $\rho_G$~\eqref{eq:app_rho_part}, namely
\begin{equation}
\begin{aligned}
    \label{eq:k-lo-duals-mse}
    \argmin_{D \in \mathcal{D}_G} {\mathcal{E}}_{\rho_{G}}(D) &=  \argmin_{D \in \mathcal{D}_G} \sum_m \Tr[\rho_G \Pi_m] \Tr[D_m^2]
    = \argmin_{D \in \mathcal{D}_G} \prod_{g \in G} \sum_{m_g}\Tr[\rho_g \Pi_{m_g}^{(g)}] \Tr[D_{m_g}^{(g)\,2}] \\
    &= \qty(\argmin_{D \in\mathcal{D}_g} \sum_{m_g}\Tr[\rho_g \Pi_{m_g}^{(g)}] \Tr[D_{m_g}^{2}]\,,~\text{for } g \in G) \\
    & = \qty(\argmin_{D \in\mathcal{D}_g} \mathcal{E}_{\rho_g}(D)\,,~\text{for } g \in G) = \qty(\hat{D}_{g} ~ \text{for } g \in G) =: D_{k\text{-LO}}\,,
\end{aligned}
\end{equation}
where in the first and third line we used the fact that $\smash{\argmin_x f(x) + c = \argmin_x f(x)}$ to first neglect the term $\smash{\Tr[\rho_G^2]}$ in $\smash{\mathcal{E}_{\rho_G}(\cdot)}$~\eqref{eq:mse_state}, and then add the terms $\Tr[\rho_g^2]$ in each of the functions to reconstruct the local MSEs $\smash{\mathcal{E}_{\rho_g}(\cdot)}$~\eqref{eq:mse_local}.

It is important to note that $\smash{\mathcal{E}_{\rho}}(\cdot)$~\eqref{eq:mse_state} and $\smash{\mathcal{E}_{\rho_G}}(\cdot)$~\eqref{eq:k-lo-duals-mse} represent different figures of merit, and thus the duals they define, $\smash{\hat{D}_G}$~\eqref{eq:k-struct-duals} and $\smash{D_{k\text{-LO}}}$~\eqref{eq:k-lo-duals-mse} will, in general, also be different. Specifically, while the former are duals that have a $k$-local structure and are optimal for the \textit{global} state $\rho$, the latter are duals that have a $k$-local structure but are instead optimal for the partitioned product state $\rho_G$~\eqref{eq:app_rho_part}. Consequently, $k$-LO duals are not guaranteed to be the best $k$-local duals for any given state $\rho$. Only when $\rho = \rho_G$, \ie when $\rho$ is in fact a $k$-local product state, are $D_{k\text{-LO}}$ guaranteed to be the truly optimal ones. This also implies that if the reduced states in $\rho_G$ end up being maximally mixed states, which is the case for, \eg\!, a global state which is maximally entangled, then $D_{k\text{-LO}}$ will be equivalent to canonical duals and not present any advantage with respect to them.

\subsection{Toy examples}
\label{sec:toy-examples}
\begin{figure*}
\subfloat[Variances of the estimation of $ZZ$ and MSE of state estimation for the mixed states $\rho = (1-q) \dyad{00} + q \dyad{11}$ with varying $q$.]{\includegraphics[width=\columnwidth]{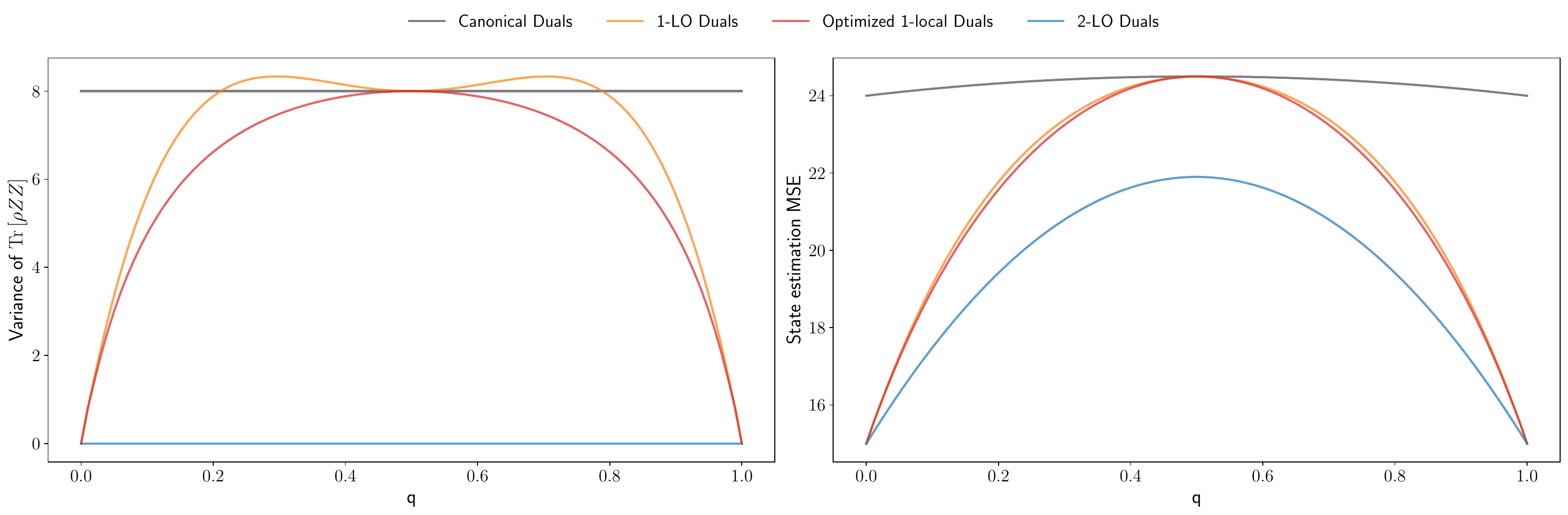}}

\subfloat[Variances of the estimation of $ZZ$ and MSE of state estimation for the pure states $\ket{\psi} = \sqrt{1-q^2} \ket{00} + q \ket{11}$ with varying $q$.]{\includegraphics[width=\columnwidth]{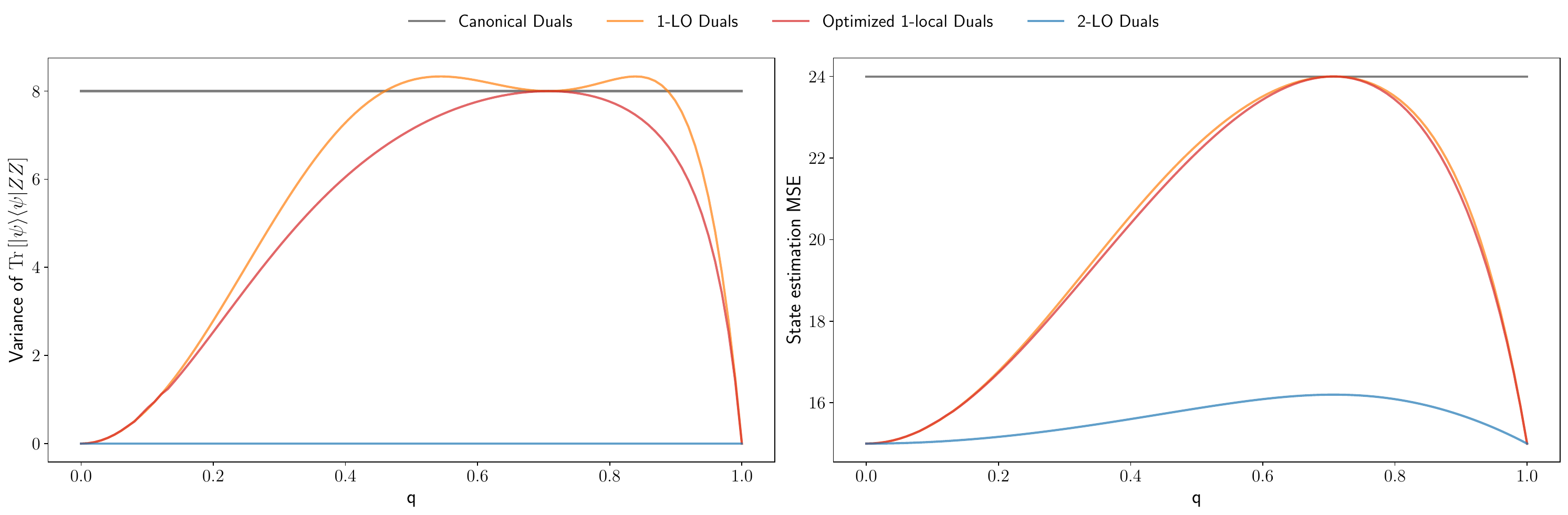}}

\caption{The estimation variance of the $ZZ$ observable~\eqref{eq:zz_var} and the state estimation MSE~\eqref{eq:mse_state} for two 2-qubit mixed and pure states, for various types of duals. For some values of $q$, the canonical estimator has a lower observable variance than the 1-LO estimator, even though the MSE error is consistently lower for the 1-LO duals. In these cases, 2-LO duals are equivalent to the globally optimal duals, and in fact proves to be the best for both observable and state estimation tasks.}
\label{fig:counterexample-variance}
\end{figure*}
We illustrate these subtleties through some toy examples on $n=2$ qubits, where we compare the different types of duals introduced above. Also, we see that $k$-LO duals can, in fact, even yield worse estimators than canonical ones in some pathological cases. Consider the following two-qubit states parameterized by a single parameter $q \in [0,1]$, a classical mixture
\begin{equation}
    \rho = (1-q) \dyad{00} + q \dyad{11}
\end{equation}
and a pure, weighted Bell state
\begin{equation}
    \ket{\psi} = \sqrt{1-q^2} \ket{00} + q \ket{11}\,,\quad\rho = \dyad{\psi}\,.
\end{equation}
For both states, and using the Pauli POVM~\eqref{eq:pauli-oc-povm}, we consider four different types of duals
\begin{itemize}
    \item Canonical duals;
    \item $1$-LO duals, defined via Eq.~\eqref{eq:k-lo-duals-mse};
    \item Optimal $1$-local duals, obtained by minimizing the global cost function in Eq.~\eqref{eq:k-struct-duals} with duals having a product structure;
    \item $2$-LO duals, obtained by minimizing the global cost function in Eq.~\eqref{eq:mse_state}. Since we only have 2 qubits, these are the true globally optimal duals.
\end{itemize}
Results from numerical simulations of these toy models are shown in Fig.~\ref{fig:counterexample-variance}, where we report the state estimation error MSE~\eqref{eq:mse_state} and the estimator variance associated to measuring the two-qubit observable $ZZ$, defined as
\begin{equation}
    \label{eq:zz_var}
    \Var{ZZ; \rho, D} = \sum_{m=1}^{36} \Tr[\rho \Pi_m] \Tr[D_m ZZ]^2 - \expval{ZZ}^2\,.
\end{equation}
for each of the four types of duals $\qty{D_m}$ described above.

First off, as one would expect, $k$-LO duals match the variance and state estimation error of the optimized $1$-local duals as well as 2-LO duals when $q=0$ and $q=1$, that is when the state is, in fact, separable.

However, the behavior changes as the states become, respectively, entangled or classically mixed. Indeed, for $0<q<1$, one can see that the 1-LO duals are not the optimal choice in the space of duals with a 1-local structure, both in terms of observable variance and state estimation error. This suggests that it is possible to find duals with $k$-local structure that produce estimators with lower variance than $k$-LO duals through, \eg optimization~\cite{Malmi2024EnhancedEstimation}.

Additionally, despite the fact that 1-LO duals consistently reduce the error of state estimation compared to canonical duals (right panels in Fig.~\eqref{fig:counterexample-variance}), they also produce observable estimators whose variance can be larger than canonical estimators for certain ranges of $q$. This result shows that, even if the state estimation error is lower with a given set of duals, there is no guarantee that the variance of the estimation of a given observable will be lower as well.

In conclusion, the goal of this analysis was to highlight the subtleties regarding the various and equally valid definitions of duals and corresponding optimality measures. In particular, although $k$-LO duals generally prove to be very effective at yielding estimators with low variance compared to other approaches (see Sec.~\ref{sec:results} in the main text), it is not possible to theoretically prove they they will, in fact, be better for any given state. On the contrary, as seen in the example above, there may be pathological cases where $k$-LO duals can in fact yield estimators which are worse than the canonical ones. On a similar note, even though all numerical results obtained in the manuscript suggest otherwise, we also suspect that there may be cases where higher locality of the duals doesn't imply better estimation performances. On a positive note, the difference between $k$-LO duals and optimal $k$-local duals is a promising direction for finding enhanced estimators. We leave a more in-depth theoretical and numerical analysis of these scenarios as a subject for future studies.

\section{Correlated groups via graph partitioning}
\label{app:partitioning-comparison}
\begin{figure}[ht]
    \centering
    \includegraphics[width=0.75\columnwidth]{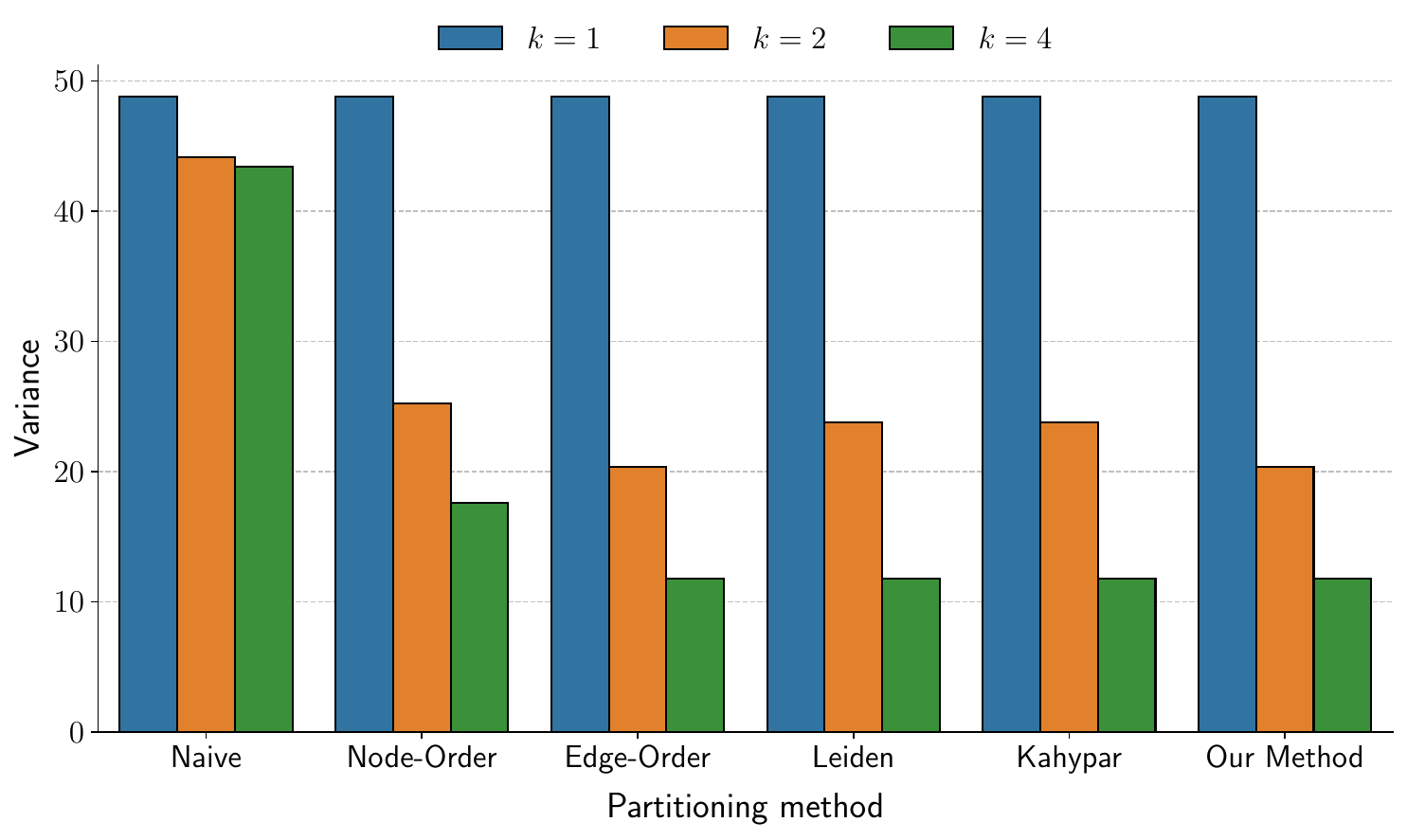}
    \caption{Comparison of partitioning methods to obtain groups of $k$ correlated qubits for different maximum group sizes for the 14-qubit H$_2$O molecule. The groupings used for this figure are shown in Table \ref{table:partitionings}. The $k$-LO duals were computed for the \textit{exact} reduced density matrices based on the partitionings.}
    \label{fig:partitioning-method-comparison}
\end{figure}

In order to obtain high-quality $k$-LO with good statistical performances, it is important to construct them so that they retain as much correlations as possible present in the measured state. As explained in Sec.~\ref{sec:graph_partitioning} in the main text, we do so by first estimating the mutual information from the experimental data, and then partitioning the qubits accordingly through a custom greedy algorithm.

However, in addition to the partitioning scheme described in the main text, other methods are possible. In this appendix we explore other approaches, including established graph-based community detections algorithms that take into account the graph of the pair-wise mutual information, and show how the statistical performances of $k$-LO duals highly depends on the strategy used.

\begin{figure}[ht!]
    \includegraphics[width=0.3\linewidth]{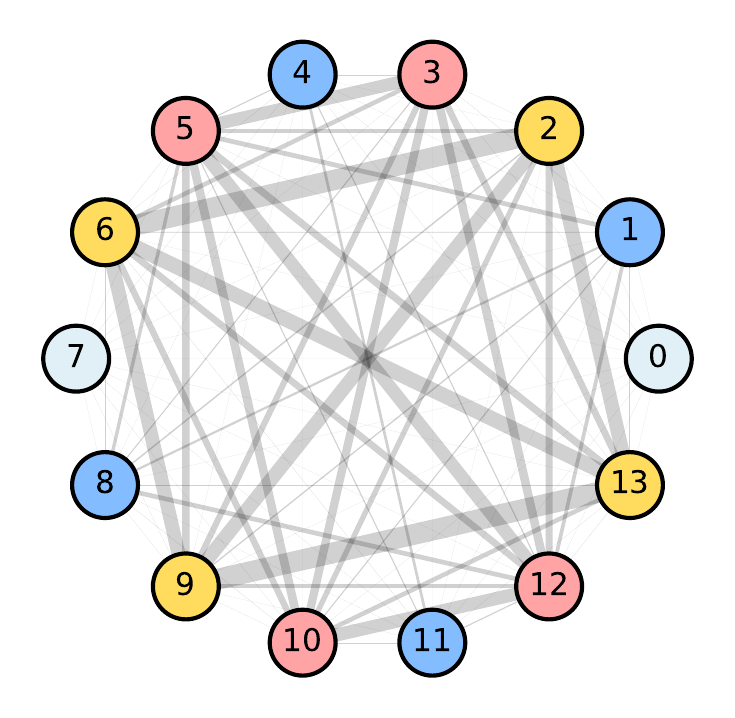}
    \caption{Mutual information graph obtained from measuring the ground state of the $\text{H}_2\text{O}$ molecule encoded on a 14-qubit state, computed from $S=10^6$ shots of the measurement \eqref{eq:pauli-oc-povm}. Edge thickness indicate the amount of correlations between qubits, and the nodes are colored according to the partitioning method described in Sec. \ref{sec:graph_partitioning}, with maximum group size set to $k=4$.}
    \label{fig:14q-mi-graph}
\end{figure}

Consider the mutual information matrix $[\text{MI}]_{ij} = I(i: j)$, where $I(i:j)$ is the mutual information of the measurement frequencies between qubit $i$ and $j$, see Eq.~\eqref{eq:cmi-freqs}. From this, one can construct a graph where each node represents a qubit, and edges between the pair-wise mutual information, and thus frame the task of finding most correlated qubits to that of graph partitioning. An example of this graph is reported in Fig.~\ref{fig:14q-mi-graph}, where we show the graph of pairwise mutual information for the 14-qubit ground-state energy of the H$_2$O molecule described in the main text.

Given the mutual information graph, we consider the following grouping algorithms:
\begin{itemize}
    \item \textit{\textit{Naive}---} The most basic method to partition the system is to \textit{not} consider the underlying mutual information graph at all, and to construct partitions just in order of the qubits. That is, for $n$ qubits and a maximum group size of $k$, one constructs $\left\lfloor N / k \right\rfloor$ groups of size $k$, and one group of size $\leq k$. The first group would simply contain the qubits with indices $\{0,1,\dots,k\}$, the second group contains $\{k+1,k+2\dots,2k\}$ and so on until the last qubit $N-1$ has been added to a group.

    This method serves as a baseline to assess the importance of using information about the correlation of the qubits when constructing the duals.

    \item \textit{\textit{Node- and Edge-order}---} To produce partitions that take into account the correlations in the state under investigation, we tested two simple greedy partition strategies, that we named \textit{node-order} or \textit{edge-order}. The former constructs groups by going through the qubits in order of their index, and adding pairs with the largest mutual information into the same group. The latter instead goes through all edges in order of their weights and constructs groups from the pairs associated to the edges. When a group has reached the maximum group size, the edge is disregarded.

    \item \textit{\textit{Community-detection}---} We also tested established general purpose graph partitioning procedures, specifically the Leiden algorithm~\cite{traag2019from} and KaHyPar~\cite{schlag2022high-quality}.
\end{itemize}

\begin{table*}
\setlength{\tabcolsep}{5pt}
\renewcommand{\arraystretch}{1.25}
\setlength{\aboverulesep}{0pt}
\setlength{\belowrulesep}{0pt}
\begin{tabular}{c|c||c}
\toprule
Method & Maximum group size & Partitioning \\
\toprule
\multirow{2}{*}{Naive} & 2 & (0, 1), (2, 3), (4, 5), (6, 7), (8, 9), (10, 11), (12, 13) \\
 & 4 & (0, 1, 2, 3), (4, 5, 6, 7), (8, 9, 10, 11), (12, 13) \\
\toprule
\multirow{2}{*}{Node-Order} & 2 & (0, 5), (1, 12), (2, 6), (3, 10), (4, 11), (7, 9), (8, 13) \\
 & 4 & (0, 5, 7, 8), (1, 9, 10, 12), (2, 3, 6, 13), (4, 11) \\
\toprule
\multirow{2}{*}{Edge-Order}  & 2 & (0, 7), (1, 8), (2, 6), (3, 10), (4, 11), (5, 12), (9, 13)  \\
 & 4 & (0, 4, 7, 11), (1, 8), (2, 6, 9, 13), (3, 5, 10, 12) \\
\toprule
\multirow{2}{*}{Leiden} & 2 & (0, 7), (1, 8), (2, 6), (3, 5), (4, 11), (9, 13), (10, 12) \\
 & 4 & (0, 4, 7, 11), (1, 8), (2, 6, 9, 13), (3, 5, 10, 12) \\
\toprule
\multirow{2}{*}{KaHyPar} & 2 & (0, 7), (1, 8), (2, 6), (3, 5), (4, 11), (9, 13), (10, 12) \\
 & 4 & (0, 1, 8), (2, 6, 9, 13), (3, 5, 10, 12), (4, 7, 11) \\
\toprule
\multirow{2}{*}{Our Method} & 2 & (0, 7), (1, 8), (2, 6), (3, 10), (4, 11), (5, 12), (9, 13) \\
 & 4 & (0, 7), (1, 4, 8, 11), (2, 6, 9, 13), (3, 5, 10, 12) \\
\hline
\end{tabular}
\caption{Partitions found using $S=10^6$ shots of the Pauli POVM on the ground state of 14-qubit H$_2$O molecules using with different methods. The underlying mutual information graph is shown in Fig.~\ref{fig:14q-mi-graph}, with node colors representing the groups found by our method.}
\label{table:partitionings}
\end{table*}

In Fig.~\ref{fig:partitioning-method-comparison} we show the variances obtained when the various partitioning methods are used to find groupings for 14-qubit H$_2$O. The partitionings themselves are shown in Table~\ref{table:partitionings}. We see that for this example, the naive approach has a substantially lower performance as other approaches, as increasing the maximum group size has a minimal effect on the variance of $k$-LO duals. We can thus conclude that, to obtain good partitionings for $k$-LO duals in general, it is necessary to use information about the underlying state.

While we note that results are highly problem-dependent, in all of our numerical simulations we have observed that KaHyPar and our method perform similarly, and out of the methods that we have studied, seem to be consistently the best both in terms of the variance of the corresponding estimator (Fig.~\ref{fig:partitioning-method-comparison}), and also in terms of modularity of the partitioned graph itself~\cite{hagberg2008exploring}.

\section{Comparison of tomography methods for obtaining the optimal frame operator}
\label{app:comparison-tomography}
\begin{figure}[ht!]
    \centering
    \includegraphics[width=0.8\columnwidth]{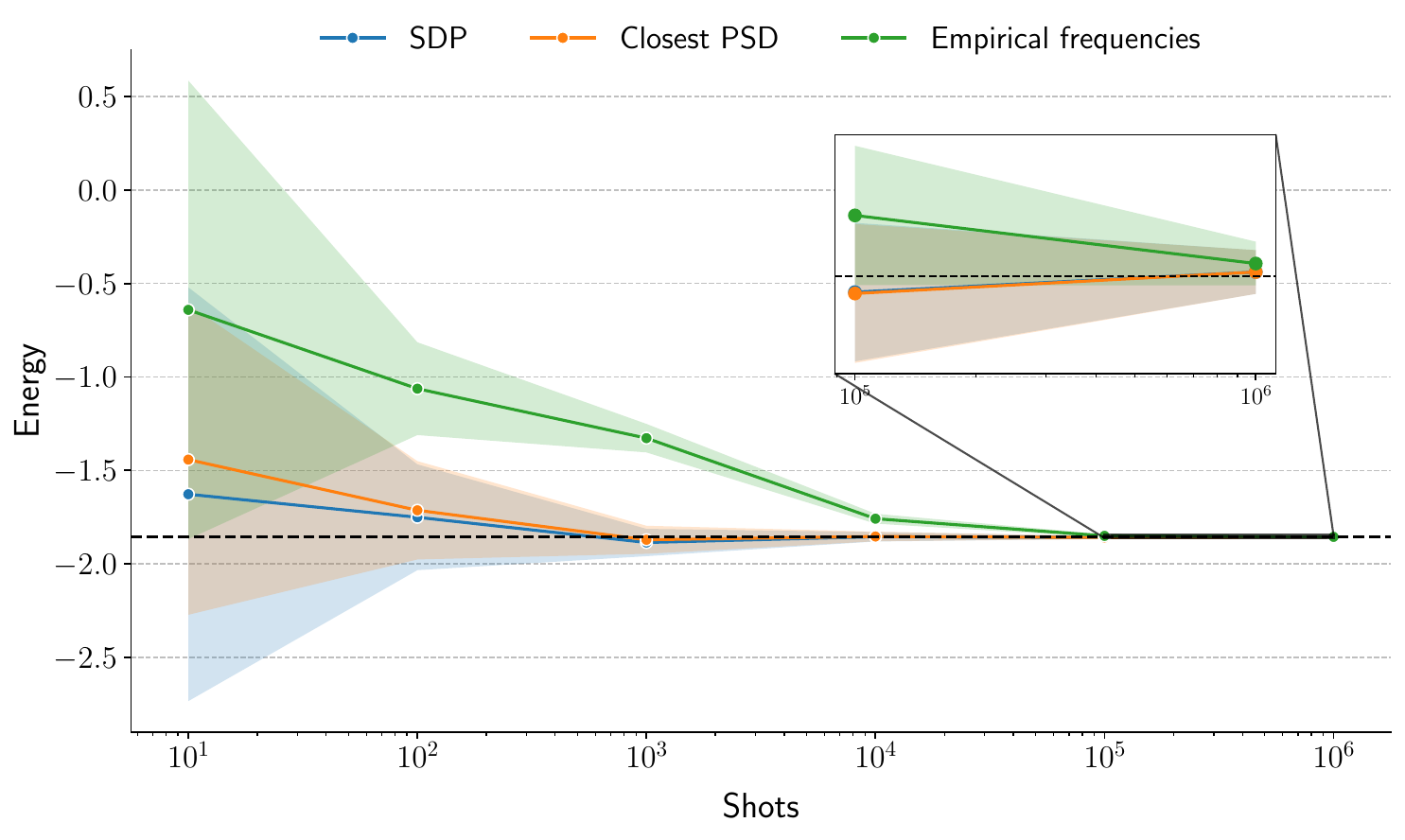}
    \caption{Comparison of methods to obtain the optimal frame operator for the ground state energy estimation of 4-qubit H$_2$. The 4-LO duals and the energy estimations were computed using the same dataset. The method with empirical frequencies uses $S_{\textrm{bias}} = 6^4 = 1296$. The energies and their errors are shown in units of $\text{Ha}$.}
    \label{fig:rdm-method-comparison}
\end{figure}

\begin{figure}[ht!]
    \centering
    \includegraphics[width=0.8\columnwidth]{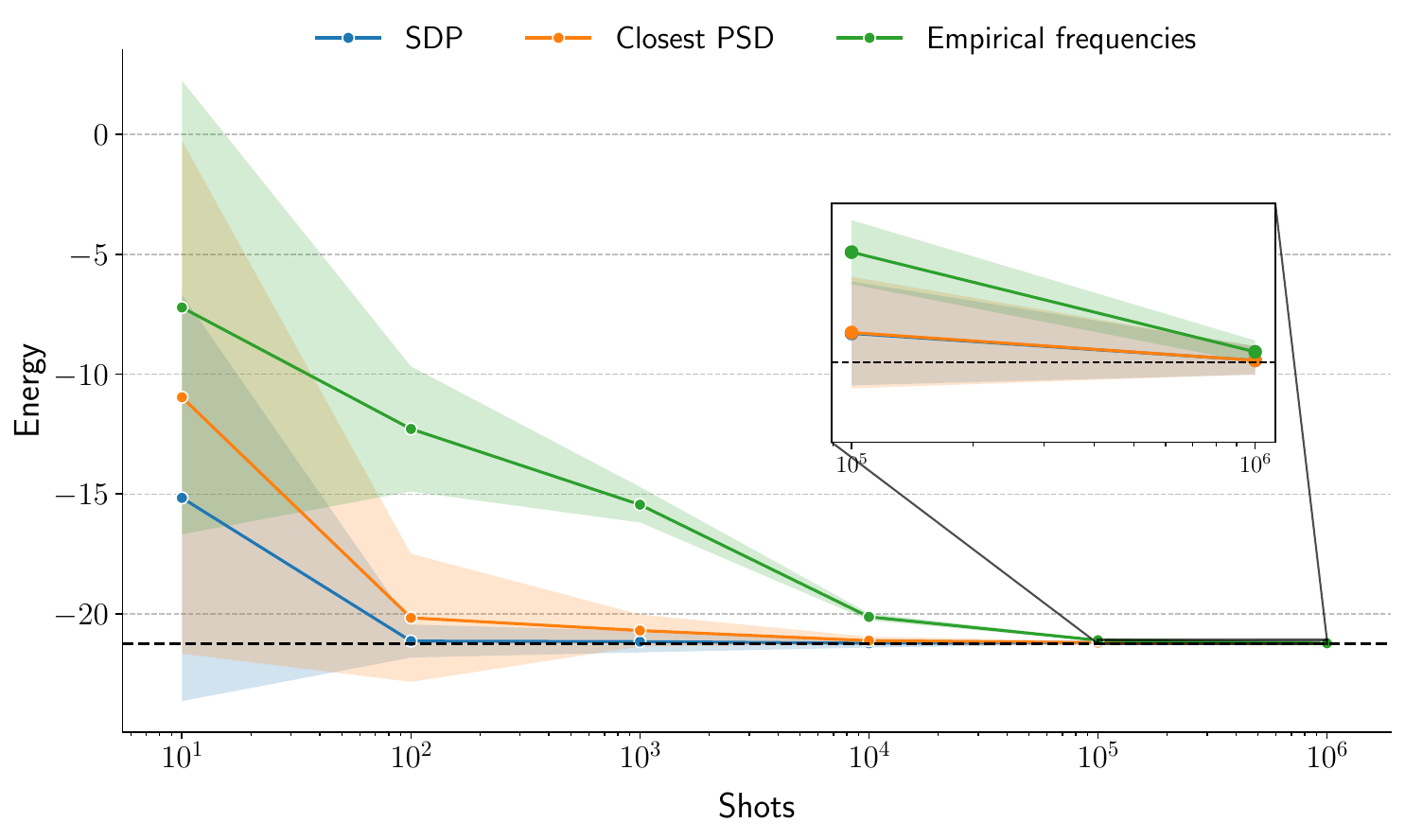}
    \caption{Comparison of methods to obtain the 4-LO frame operator for the ground state energy estimation of 16-qubit NH$_3$. The 4-LO duals and the energy estimations were computed using the same dataset. The method with empirical frequencies uses $S_{\textrm{bias}} = 6^4 = 1296$. The energies and their errors are shown in units of $\text{Ha}$.}
    \label{fig:rdm-method-comparison-nh3-same}
\end{figure}

\begin{figure}[ht!]
    \centering
    \includegraphics[width=0.8\columnwidth]{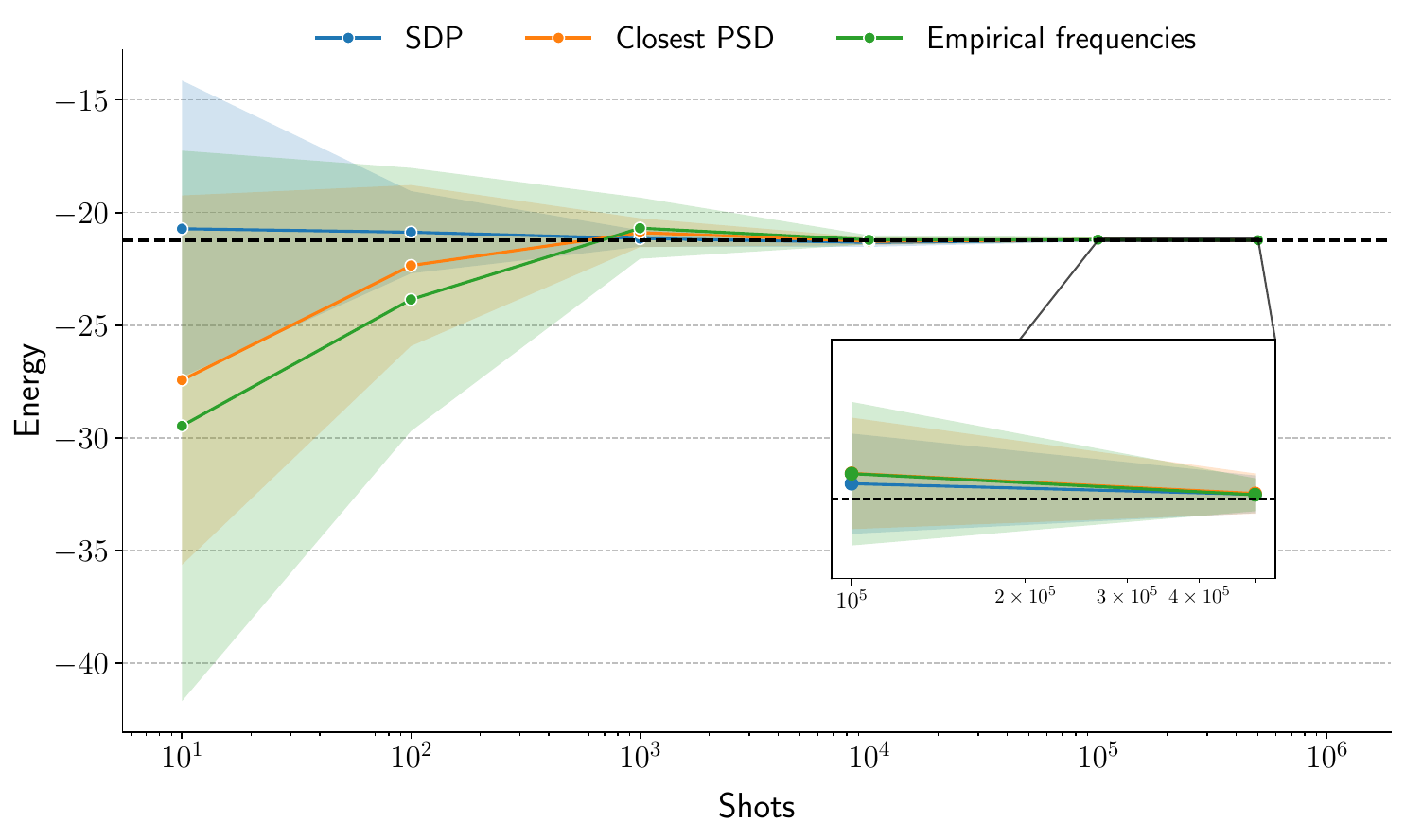}
    \caption{Comparison of methods to obtain the 4-LO frame operator for the ground state energy estimation of 16-qubit NH$_3$. The 4-LO duals and the energy estimations were computed using independent datasets. The method with empirical frequencies uses $S_{\textrm{bias}} = 6^4 = 1296$. The energies and their errors are shown in units of $\text{Ha}$.}
    \label{fig:rdm-method-comparison-nh3-independent}
\end{figure}

The final step in the construction of $k$-LO duals is to construct the $k$-local state-dependent frame operators as described in the main text \ref{sec:approx_local_states}.
To construct such frame operator, we need the probabilities $\Tr\left[\Pi_m \rho\right]$ for POVM effects $\Pi_m$ and state $\rho$.
However, we do not have access to these exact probabilities in general often due to finite statistics.
The issue is then, which values to use in place of the probabilities, such that the frame operator is similar to the exact state-dependent frame operator.
One approach is to directly use the empirical frequencies obtained from experiments as explored in Refs.~\cite{ZhouOCPOVM2014, Fischer2024DualOptimization}.
However, in practical cases with a large number of qubits, it is expected for most outcomes to never appear and as a result get associated a frequency of zero, which is problematic, as the frame operator is constructed with these frequencies in the denominator.
A solution to this issue proposed in Ref.~\cite{Fischer2024DualOptimization} is to use a constant $S_{\text{bias}}$ added to the empirical frequencies, essentially mixing them with uniform frequencies.
Note that this approach does not produce invalid duals, that would produce biased estimators, instead we would simply find duals that are optimal for another state, the one built by the empirical frequencies and the maximally mixed state, but are still valid for any state and observable.

Another approach is to perform state tomography using the empirical frequencies from the experiment, to get a better approximation of the state in question and compute the full probability distribution, given that the number of qubits is limited, as is needed for $k$-LO duals.
Given empirical frequencies $\{f_m\}_m$, one can construct an unphysical approximation of the state,
\begin{equation}
    \bar{\rho} = \sum_m{f_m D_m},
\end{equation}
where $D_m$ is the canonical dual of POVM effect $\Pi_m$,
and then finding the closest positive semi-definite (PSD) matrix in the Frobenius norm \cite{higham1988computing}.
Another approach is to perform semi-definite programming (SDP) to find the closest physical state, that would produce the distribution $\{f_m\}_m$, as discussed in Ref.~\cite{cattaneo2023self-consistent}.

In Fig.~\ref{fig:rdm-method-comparison} we compare these three approaches to obtain 4-LO duals on the ground state of 4-qubit H$_2$ for various number of shots.
For a given number of shots, we construct the frame operator, and perform the energy estimation on the same dataset.
We also show the confidence interval of 99.7\% ($3\times$standard error) of the estimation for each of the approaches.
From the results, we notice that both the SDP and closest PSD approaches produce estimations that contain the exact ground state energy within error bars.
Instead, for the approach with empirical frequencies using $S_{\text{bias}} = 1296$, estimations for number of shots between $10^2$ and $10^4$ are much further away from the exact ground state energy and do not contain it within error bars.
We suspect that these duals are, in fact, valid and produce and unbiased estimators, however when we perform the estimation using the same dataset as we have constructed the frame operator, the estimates do not seem to contain the true value within error bars. This is most likely the consequence of a type of overfitting. The exact ground state energy is within error bars only when the number of shots is either very small, so $S_{\text{bias}}$ dominates and we produce the canonical estimator, or large enough so that the empirical frequencies approximate the exact probability distribution well.

When the duals depend on the measurement data, the estimator
\begin{align}
    \bar{o} = \sum_m{f_m \Tr[D_m(\{f\}) O]}
\end{align}
is generally biased due to statistical dependence between the frequencies $f_m$ and the duals $D_m(\{f\})$.
In the empirical-frequencies approach of Ref.~\cite{Fischer2024DualOptimization}, the duals depend on each $f_m$ through $1/f_m$ terms in the frame operator, which may even lead to an amplification of the correlation between the frequencies and duals --- particularly for unlikely outcomes where $f_m$ is small --- causing estimator bias at moderate shot counts or low values of $S_{\text{bias}}$.
In contrast, our approach first constructs a tomographic state from the measurement data, mapping the $d^k$-dimensional frequency space (with $d$ being the number of measurement outcomes per qubit), through the $4^k$-dimensional operator space, and only then computes the probabilities entering the frame operator from this reconstructed state.
We conjecture that this dimensional bottleneck strongly attenuates the coupling between individual frequencies and the resulting duals, suppressing the bias to negligible levels.

This is consistent with our numerical observations, where no detectable bias is found across all experiments performed.
This effect is illustrated in Figs.~\ref{fig:rdm-method-comparison-nh3-same} and \ref{fig:rdm-method-comparison-nh3-independent}, where we show additional numerical results for the 16-qubit NH$_3$ molecule.
In Fig.~\ref{fig:rdm-method-comparison-nh3-same}, we show that the empirical-frequencies method exhibits bias at intermediate-shot regimes while the closest-PSD and SDP methods do not, similarly to the results shown in Fig.~\ref{fig:rdm-method-comparison}.
On the other hand, if any residual bias is of concern, one may use an independent dataset to produce the duals, ensuring with certainty that the estimator will be unbiased.
In Fig.~\ref{fig:rdm-method-comparison-nh3-independent}, we show results where independent datasets were used for the dual construction and the estimation.
As expected, all methods now exhibit unbiasedness; however the empirical-frequencies approach produces the estimator with the highest uncertainty while the SDP approach has the smallest error bars.
This confirms that the advantage of our approach is twofold: it suppresses the bias from data reuse, and it yields higher-quality duals due to a more effective state reconstruction. We leave a rigorous theoretical analysis of this bias suppression as a subject for future work.

\section{Comparison of measurement methods for molecular ground states}
\label{app:variance-table}
We present numerical values for the comparison of variances between different measurement schemes in Table \ref{table:molecular-variances}.
The values correspond to the results shown in Fig.~\ref{fig:main-image}.

\begin{table*}[ht]
\setlength{\tabcolsep}{5pt}
\renewcommand{\arraystretch}{1.1}
\setlength{\aboverulesep}{0pt}
\setlength{\belowrulesep}{0pt}
\begin{tabular}{l|cccc||>{\bfseries}c>{\bfseries}c>{\bfseries}c}
\toprule
\multirow{2}{*}{Molecule} & \multirow{2}{*}{CS-Pauli} & \multirow{2}{*}{LBCS} & \multirow{2}{*}{GBCS} & \multirow{2}{*}{OGM} & \multicolumn{3}{c}{$k$-LO Duals} \\
 &  &  &  &  & \normalfont{1} & \normalfont{2} & \normalfont{4} \\
\hline
H$_2$ (4) & 1.97 & 1.86 & 0.36 & 0.42 & 0.80 & 0.71 & 0.67 \\
H$_2$ (8) & 51.4 & 17.7 & 6.2 & 5.51 & 3.42 & 3.01 & 2.95 \\
LiH (12) & 266 & 14.8 & 8.5 & 3.09 & 1.02 & 0.81 & 0.79 \\
BeH$_2$ (14) & 1670 & 67.6 & 32.8 & 15.44 & 38.61 & 6.68 & 6.32 \\
H$_2$O (14) & 2840 & 258 & 294.4 & 39.64 & 48.72 & 20.65 & 13.86 \\
NH$_3$ (16) & 14396 & 353 & - & - & 898 & 157 & 41 \\
\hline
\end{tabular}
\caption{Variances for different molecules and measurement schemes in units of $\text{Ha}^2$.}
\label{table:molecular-variances}
\end{table*}

In Table~\ref{table:molecular-rmse}, we report the root mean square error (RMSE) of different methods, which quantifies the average error over multiple independent energy estimations. Specifically, the RMSE is defined as
\begin{equation}
    \label{eq:rmse}
    \text{RMSE} = \sqrt{\frac{1}{R} \sum_{r=1}^R \qty(\bar{o}_r - \Tr[\rho O])^2}\,,
\end{equation}
where $\bar{o}_r$ is the energy estimated in the $r$-th experiment out of $R=1000$ independent realizations, each using $S=10^3$ sampled shots. The $k$-LO duals used for these estimations are the same as those employed for the variance calculations above, obtained from a separate dataset of $S=10^6$ shots. We do not use $k$-LO duals derived from the same $S=10^3$ shots as the RMSE itself because this number of shots renders a poor partial state tomography~\eqref{eq:sdp} required for high-locality $k$-LO duals. We note, however, that a separate dataset is not required to compute the duals: empirically, we observe that reusing the same measurement data for both dual construction and estimation yields estimators without detectable bias (see App.~\ref{app:comparison-tomography}). Alternatively, one may construct the $k$-LO duals from a classical approximation of the $k$-local RDMs.

We note that the RMSE reported here is primarily for comparison with other methods and not as a realistic benchmark of measurement performance. In practice, it would be more interesting to assess the measurement overhead to reach chemical precision ($1.6 \times 10^{-3} \text{Ha}$) at a regime beyond $S=10^3$ shots, which would also improve the quality of the local tomography required for $k$-LO duals. A more thorough benchmark at a higher-shot regime would be a very informative major undertaking, and we leave it as a subject of future work.

\begin{table*}[!ht]
\setlength{\tabcolsep}{5pt}
\renewcommand{\arraystretch}{1.1}
\setlength{\aboverulesep}{0pt}
\setlength{\belowrulesep}{0pt}
\begin{tabular}{l|ccccc||>{\bfseries}c>{\bfseries}c>{\bfseries}c}
\toprule
\multirow{2}{*}{Molecule} & \multirow{2}{*}{CS-Pauli~\cite{HuangShadows2020}} & \multirow{2}{*}{LBCS~\cite{HadfieldLBCS2022}}  & \multirow{2}{*}{OGM~\cite{WuOGM2023}} & \multirow{2}{*}{Derand~\cite{huang2021efficient}} & \multirow{2}{*}{AEQuO~\cite{shlosberg2023adaptive}} & \multicolumn{3}{c}{$k$-LO Duals} \\
 &  &  &  &  &  & \normalfont{1} & \normalfont{2} & \normalfont{4} \\
\hline
H$_2$ (4) & 0.048 & 0.043 & 0.011 & 0.010 & - & 0.029 & 0.027 & 0.027 \\
H$_2$ (8) & 0.223 & 0.128 & 0.051 & 0.067 & 0.025 & 0.060 & 0.058 & 0.058 \\
LiH (12) & 0.548 & 0.122 & 0.036 & 0.063 & 0.020 & 0.032 & 0.029 & 0.028 \\
BeH$_2$ (14) & 2.936 & 0.275 & 0.072 & 0.103 & 0.037 & 0.107 & 0.093 & 0.080 \\
H$_2$O (14) & 1.328 & 0.549 & 0.129 & 0.257 & 0.062 & 0.167 & 0.151 & 0.119 \\
NH$_3$ (16) & 2.298 & 0.484 & 0.151 & 0.225 & - & 0.353 & 0.247 & 0.148 \\
\toprule
\end{tabular}
\caption{RMSE~\eqref{eq:rmse} for different molecules and measurement schemes in units of $\text{Ha}$. (see Sec.~\ref{ssec:comparison} for details on each of them), computed with $R=1000$ independent estimations, each obtained with $S=10^3$ shots. The values for the other methods are taken from the relevant sources.}
\label{table:molecular-rmse}
\end{table*}

\section{Two-point correlation functions on a spin system}
\label{app:results-ising}

\begin{figure}[ht]
    \centering
    \includegraphics[width=0.75\columnwidth]{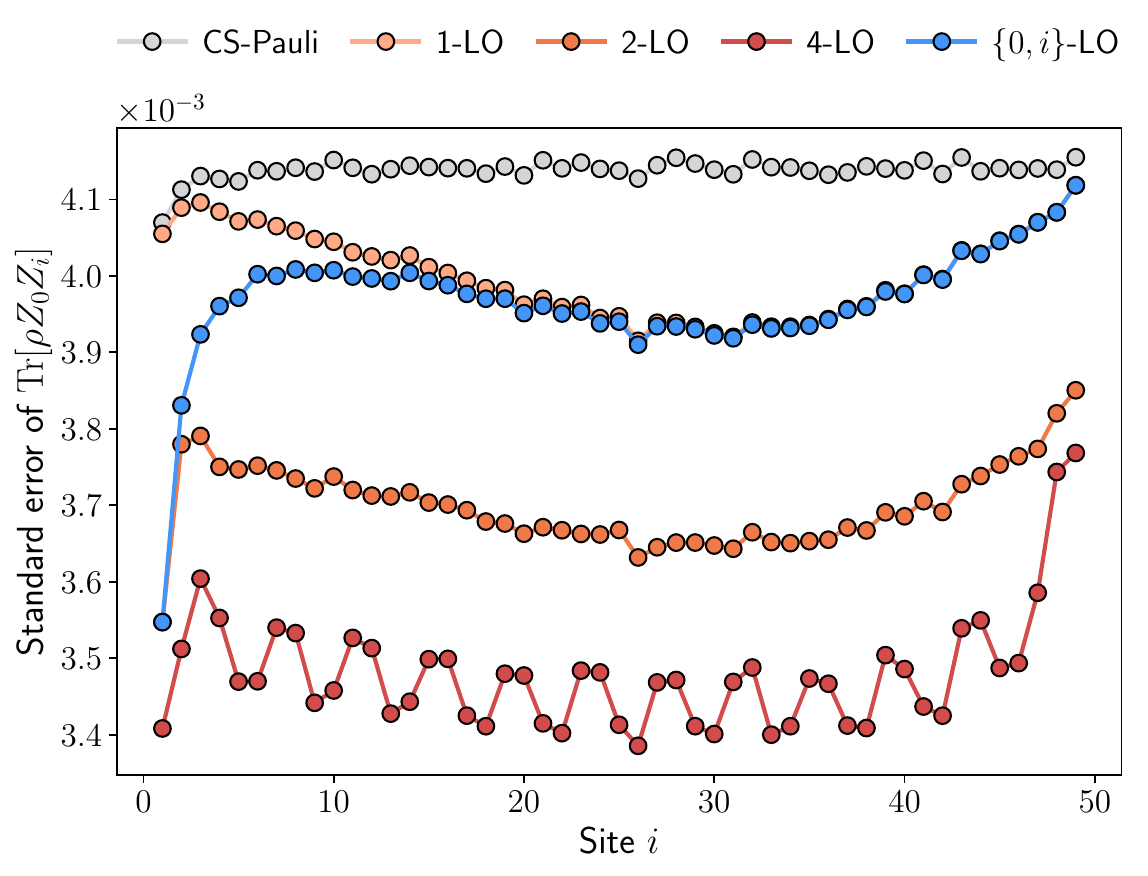}
    \caption{The standard error $\sqrt{\Var{\bar{o}}/S}$ ($\sim 68\%$ CI) of the estimation of two-point correlation functions for 1D TFIM using $2^{19}$ shots from random Pauli measurements with canonical~\cite{HuangShadows2020} (CS-Pauli) and $k$-LO duals. $\{0,i\}$-LO duals corresponds to the situation where we first trace out all but systems $0$ and $i$, and then compute the optimal duals for this subsystem.}
    \label{fig:tfim-variance}
\end{figure}

It is known that classical shadows based on Pauli measurements excel at estimating local observables~\cite{HuangShadows2020}. We study the effect of employing a more informed choice of duals on two-point correlation functions of the one-dimensional anti-ferromagnetic transverse field Ising model (TFIM) on $n=50$ qubits, which is used as a test case in Refs.~\cite{carrasquilla2019reconstructing, HuangShadows2020}. The tensor network representation of the ground state of such a model is available at Ref.~\cite{carrasquilla2019github}. In Fig.~\ref{fig:tfim-variance}, we report results for the two-point correlation observables $O = Z_0\otimes Z_i$ between the first and $i$-th qubit. As expected, we see that $k$-LO duals always outperform the canonical duals in terms of estimation accuracy.
We note that if one is interested in only estimating the $ZZ$ correlation functions, the optimal strategy would be to directly measure in the $Z$ basis, naturally at the cost of not being able to estimate other observables in pure post-processing.

Interestingly, we also notice something seemingly counter-intuitive. For most of the correlation functions, $2$- or $4$-LO duals defined on the global state are more precise than the ones constructed considering only the information at the relevant qubits $0$ and $i$, obtained by marginalizing the measurements outcomes to these qubits only, reported as $\qty{0, i}$-LO in the figure. As we move through the chain, the variance of these $\{0,i\}$-LO duals then converges to that of $1$-LO duals, which can be understood by the fact that correlations between sites are suppressed when the two are far from each other, thus leaving no room for improvements for 2-local duals over 1-local tensor product ones. This is not the case of $2$- and $4$-LO duals which, by encoding global information on the state, are capable of providing more precise estimators. This highlights the seemingly counter-intuitive fact that even the estimations of local observables can benefit from a correlated post-processing scheme.

\end{document}